\documentclass[aos,preprint]{imsart}

\usepackage[OT1]{fontenc}
\usepackage{amsthm,amsmath,amssymb,natbib}
\usepackage[bookmarks=false]{hyperref}
\usepackage{hypernat}

\usepackage{mdwlist}
\usepackage{graphicx}
\usepackage{color}
\usepackage{epsfig}
\usepackage{subfig}
\usepackage{afterpage}
\usepackage{cancel}

\usepackage{algorithm}
\usepackage{algpseudocode}

\setattribute{journal}{name}{}

\startlocaldefs

\numberwithin{algorithm}{section}
\numberwithin{equation}{section}

\renewcommand{\algorithmicrequire}{\textbf{Outputs:}}
\renewcommand{\algorithmicensure}{\textbf{Inputs:}\hspace{0.28cm}}

\def\mathbbi#1{\textbf{#1}}

\newcommand{\distUni}{\textsf{Un}}
\newcommand{\distNorm}{\mathcal{N}}

\newcommand{\distGP}{\mathcal{GP}}

\newcommand{\mcI}{\ensuremath{\mathcal{I}}}
\newcommand{\mcC}{\ensuremath{\mathcal{C}}}
\newcommand{\mcX}{\ensuremath{\mathcal{X}}}

\newcommand{\mcG}{\ensuremath{\mathcal{G}}}

\newcommand{\mcM}{\ensuremath{\mathcal{M}}}
\newcommand{\mcZ}{\ensuremath{\mathcal{Z}}}

\newcommand{\mcW}{\ensuremath{\mathcal{W}}}
\newcommand{\mcD}{\ensuremath{\mathcal{D}}}

\newcommand{\reals}{\ensuremath{\mathbb{R}}}
\newcommand{\bhg}{\ensuremath{\boldsymbol{\hat{g}}}}
\newcommand{\bg}{\ensuremath{\boldsymbol{g}}}

\newcommand{\biX}{\ensuremath{\mathbbi{X}}}
\newcommand{\biG}{\ensuremath{\mathbbi{G}}}
\newcommand{\bhiX}{\ensuremath{\hat{\mathbbi{X}}}}
\newcommand{\bhiG}{\ensuremath{\hat{\mathbbi{G}}}}
\newcommand{\half}{\ensuremath{\frac{1}{2}}}
\newcommand{\given}{\ensuremath{\,|\,}}

\newcommand{\hM}{\ensuremath{\hat{M}}}

\newcommand{\naturals}{\ensuremath{\mathbb{N}}}
\newcommand{\hmcM}{\ensuremath{\hat{\mathcal{M}}}}
\newcommand{\hmcG}{\ensuremath{\hat{\mathcal{G}}}}

\newcommand{\hg}{\ensuremath{\hat{g}}}

\newcommand{\htheta}{\ensuremath{\hat{\theta}}}
\newcommand{\hpsi}{\ensuremath{\hat{\psi}}}
\newcommand{\bx}{\ensuremath{\boldsymbol{x}}}
\newcommand{\btx}{\ensuremath{\boldsymbol{\tilde{x}}}}
\newcommand{\bhx}{\ensuremath{\boldsymbol{\hat{x}}}}
\newcommand{\bw}{\ensuremath{\boldsymbol{w}}}
\newcommand{\btw}{\ensuremath{\boldsymbol{\tilde{w}}}}
\endlocaldefs

\begin{document}

\begin{frontmatter}
  
  \title{Nonparametric Bayesian Density Modeling with Gaussian Processes}
  \runtitle{Gaussian Process Density Modeling}

    \begin{aug}
    \author{\fnms{Ryan P.} \snm{Adams}\thanksref{t1}\corref{}%
      \ead[label=u1,url]{http://www.cs.toronto.edu/~rpa/}%
      \ead[label=e1]{rpa@cs.toronto.edu}},
    \author{\fnms{Iain} \snm{Murray}\ead[label=e2]{murray@cs.toronto.edu}}
    \and
    \author{\fnms{David J.C.}
      \snm{MacKay}\ead[label=e3]{mackay@mrao.cam.ac.uk}}
    \runauthor{R.P.\ Adams et al.}

    \affiliation{University of Toronto and University of Cambridge}
    \address{Ryan P. Adams, Iain Murray\\
      Department of Computer Science\\
      University of Toronto\\
      10 King's College Road\\
      Toronto, Ontario M5S 3G4, CA\\
      \printead{e1}\\      
      \phantom{E-mail:\ }\printead*{e2}}
    \address{David J.C. MacKay\\
      Cavendish Laboratory\\
      University of Cambridge\\
      Madingley Road\\
      Cambridge CB3 0HE, UK\\
      \phantom{Email:\ }\printead*{e3}}   
   
    \thankstext{t1}{Supported by the Canadian Institute for Advanced
      Research.}
    \end{aug}

    \begin{abstract}
      We present the \textit{Gaussian process density sampler} (GPDS), an
      exchangeable generative model for use in nonparametric Bayesian
      density estimation.  Samples drawn from the GPDS are consistent with
      exact, independent samples from a distribution defined by a
      density that is a transformation of a function drawn from a
      Gaussian process prior. Our formulation allows us to infer an unknown
      density from data using Markov chain Monte Carlo, which gives samples
      from the posterior distribution over density functions and from the
      predictive distribution on data space.  We describe two such MCMC
      methods.  Both methods also allow inference of the hyperparameters of
      the Gaussian process.
    \end{abstract}


    
 \end{frontmatter}


  \section{Introduction}
  We propose a method for incorporating a Gaussian process into a
  prior on probability density functions.  While such constructions
  have been proposed before
  \citep{leonard-1978a,thorburn-1986a,lenk-1988a,lenk-1991a,csato-2002a,tokdar-ghosh-2007a,tokdar-2007a},
  ours is the first that allows a procedure for drawing exact and
  exchangeable data samples from a density drawn from the prior.  We
  call this prior and the associated procedure the \textit{Gaussian
    process density sampler} (GPDS).  Given data, this generative
  prior allows us to perform inference of the unnormalised density.
  We present two Markov chain Monte Carlo (MCMC) algorithms for
  performing this inference, one based on exchange sampling
  \citep{murray-etal-2006a} and the other based on inferring the
  latent generative history.  In both cases we are also able to infer
  the parameters governing the covariance kernel, and draw samples
  from the predictive distribution on data space.

  Bayesian nonparametric inference is appealing because it allows models to
  include an arbitrary number of parameters, without requiring expensive
  dimensionality-altering computations for inference.  The most popular
  tool for nonparametric Bayesian modeling of an unknown probability
  measure is the Dirichlet process \citep{ferguson-1973a} and related
  constructions (e.g., \citet{pitman-yor-1997a} and
  \citet{ishwaran-james-2001a}).  Samples from the Dirichlet process,
  however, are discrete distributions with probability one.  For many
  inference problems, we wish to model probabilities on continuous spaces
  and in such problems our prior beliefs are often best captured by
  a distribution over probability density functions.

  To fill the gap between nonparametric priors on discrete distributions
  and nonparametric priors on continuous densities, the Dirichlet process
  is frequently used to add a countably-infinite number of parameters into
  a continuous model.  The most popular example is the infinite mixture of
  parametric distributions \citep{escobar-west-1995a}, another example is
  kernel convolution \citep{lo-1984a}. The Dirichlet diffusion tree
  \citep{neal-2001a,neal-2003a} and  P{\'o}lya trees
  \citep{lavine-1992a,lavine-1994a} provide more direct nonparametric
  priors on distributions and, in contrast to the Dirichlet process, can
  produce densities.
  All of these priors are based on beliefs of an underlying structure,
  either a clustering or tree-based hierarchy.

  Prior beliefs about a distribution over data are sometimes best expressed
  directly in terms of the
  probability density function --- its continuity, support and smoothness
  properties, for example.  There is a rich literature on incorporating
  prior beliefs about functions into nonparametric Bayesian regression
  models, using splines, neural networks and stochastic processes (e.g.,
  \citet{dimatteo-etal-2001a}, \citet{mackay-1992a},
  and~\citet{ohagan-1978a}).  However, priors on general functions have
  largely resisted application to density estimation, due to the
  requirements that probability density functions be nonnegative and
  integrate to one.  This work introduces the first fully-nonparametric
  Bayesian kernel method for density estimation that does not require a
  finite-dimensional approximation to perform inference.

  \section{The Gaussian process density sampler prior}
  \label{sec:density-prior}
  The GPDS provides a probability distribution on a space~$\mcX$, which we
  call the \textit{data space}.  In many problems,~$\mcX$ is
  the~$D$-dimensional real space~$\reals^{D}$.

  We first place a Gaussian process prior over a scalar
  function~${g(\bx):\mcX\to\reals}$.  This means that the prior distribution
  over any discrete set of function values, $\{g(\bx_n)\}_{n=1}^N$, is a
  multivariate normal distribution.  These distributions can be
  consistently defined with a positive definite covariance
  function~${C(\cdot,\cdot):\mcX\times\mcX\rightarrow\reals}$ and a mean
  function~${m(\cdot):\mcX\rightarrow\reals}$.  The mean and covariance
  functions are parameterised by \textit{hyperparameters}~$\theta$. For a
  more detailed review of Gaussian processes see, e.g.,
  \citet{rasmussen-williams-2006a}.

\begin{figure}[t]
    \centering
    \subfloat[][{\scriptsize$\ell_x\!=\!1$, $\ell_y\!=\!1$,
      $\alpha\!=\!1$}]{%
      \centering%
      \includegraphics[width=0.49\textwidth]%
      {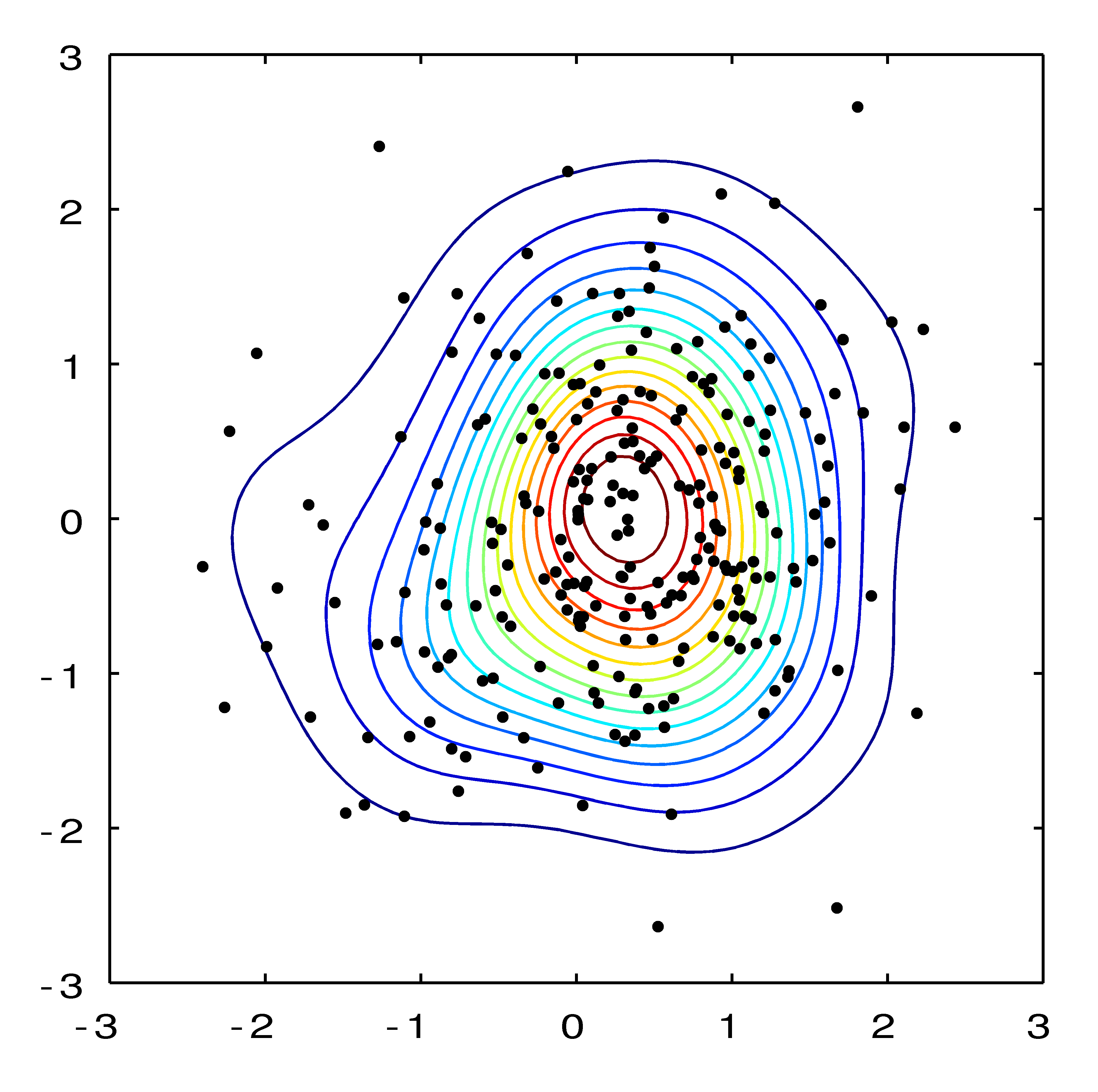}%
    }~\quad~%
    \subfloat[][{\scriptsize$\ell_x\!=\!1$, $\ell_y\!=\!1$,
        $\alpha\!=\!5$}]{%
      \centering%
      \includegraphics[width=0.49\textwidth]%
      {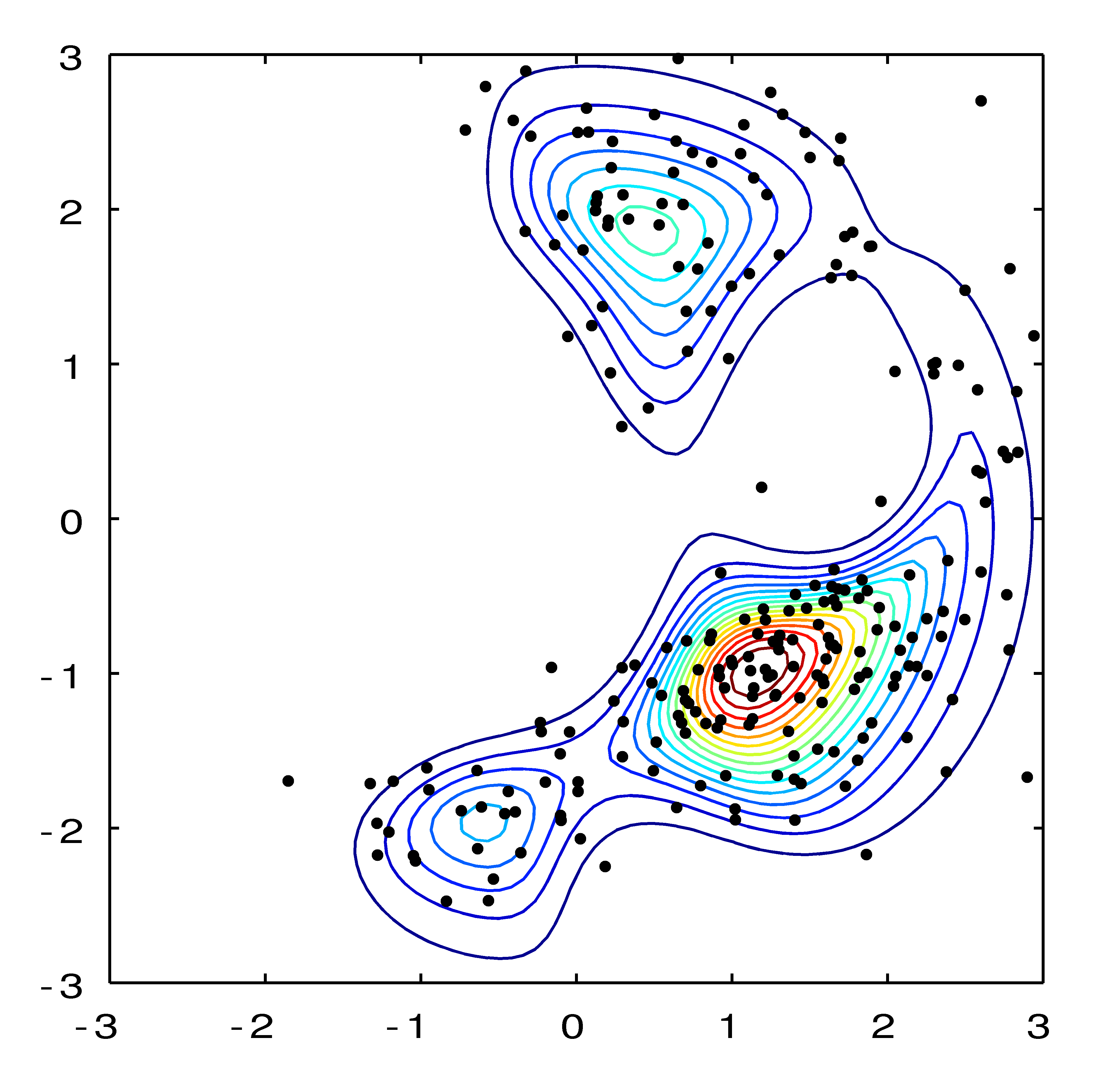}%
    }\\%
    \subfloat[][{\scriptsize$\ell_x\!=\!\frac{1}{4}$,
        $\ell_y\!=\!\frac{1}{4}$, $\alpha\!=\!1$}]{%
      \centering%
      \includegraphics[width=0.49\textwidth]%
      {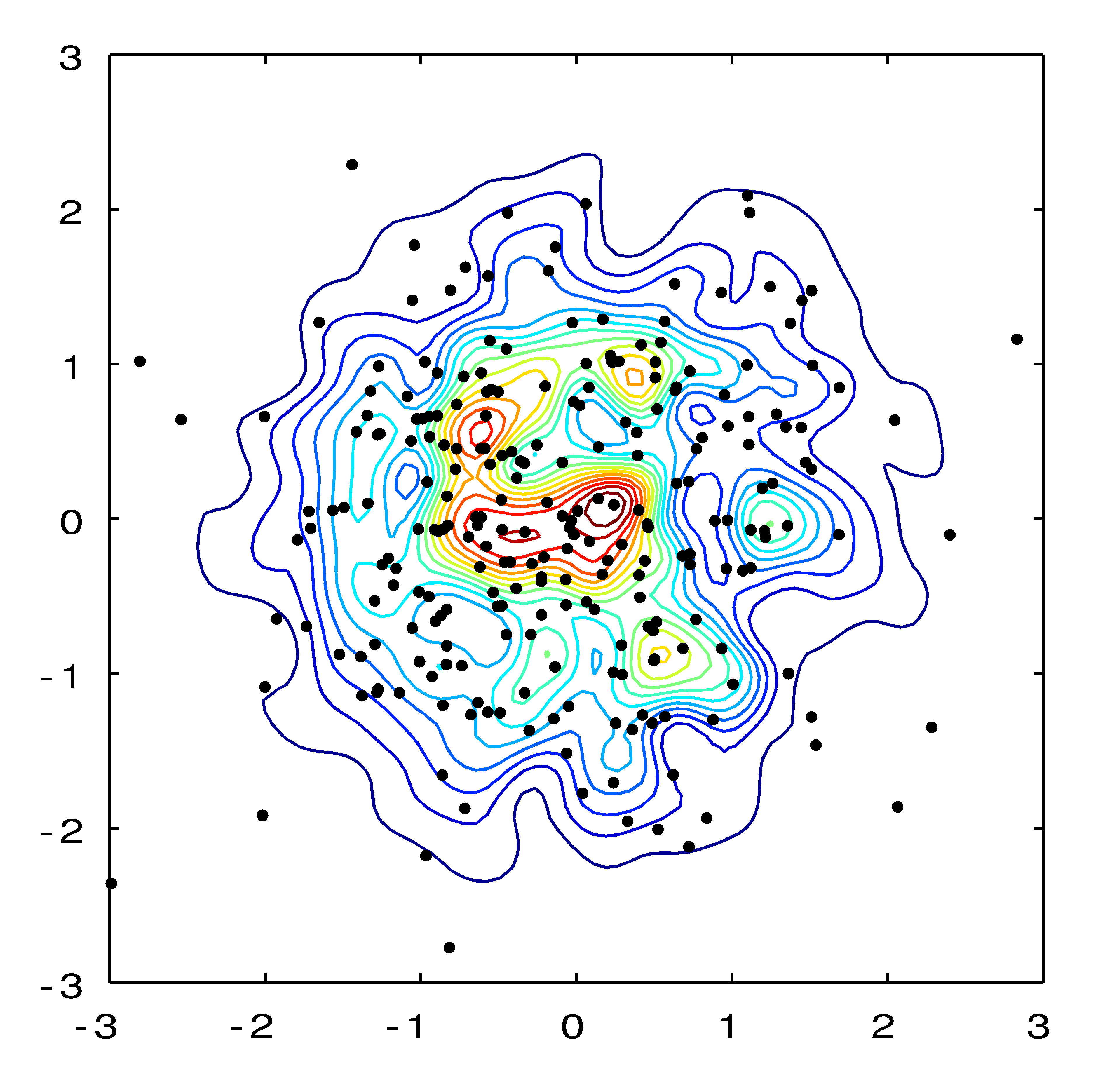}%
    }~\quad~%
    \subfloat[][{\scriptsize$\ell_x\!=\!1$, $\ell_y\!=\!\frac{1}{4}$,
      $\alpha\!=\!3$}]{%
      \centering%
      \includegraphics[width=0.49\textwidth]%
      {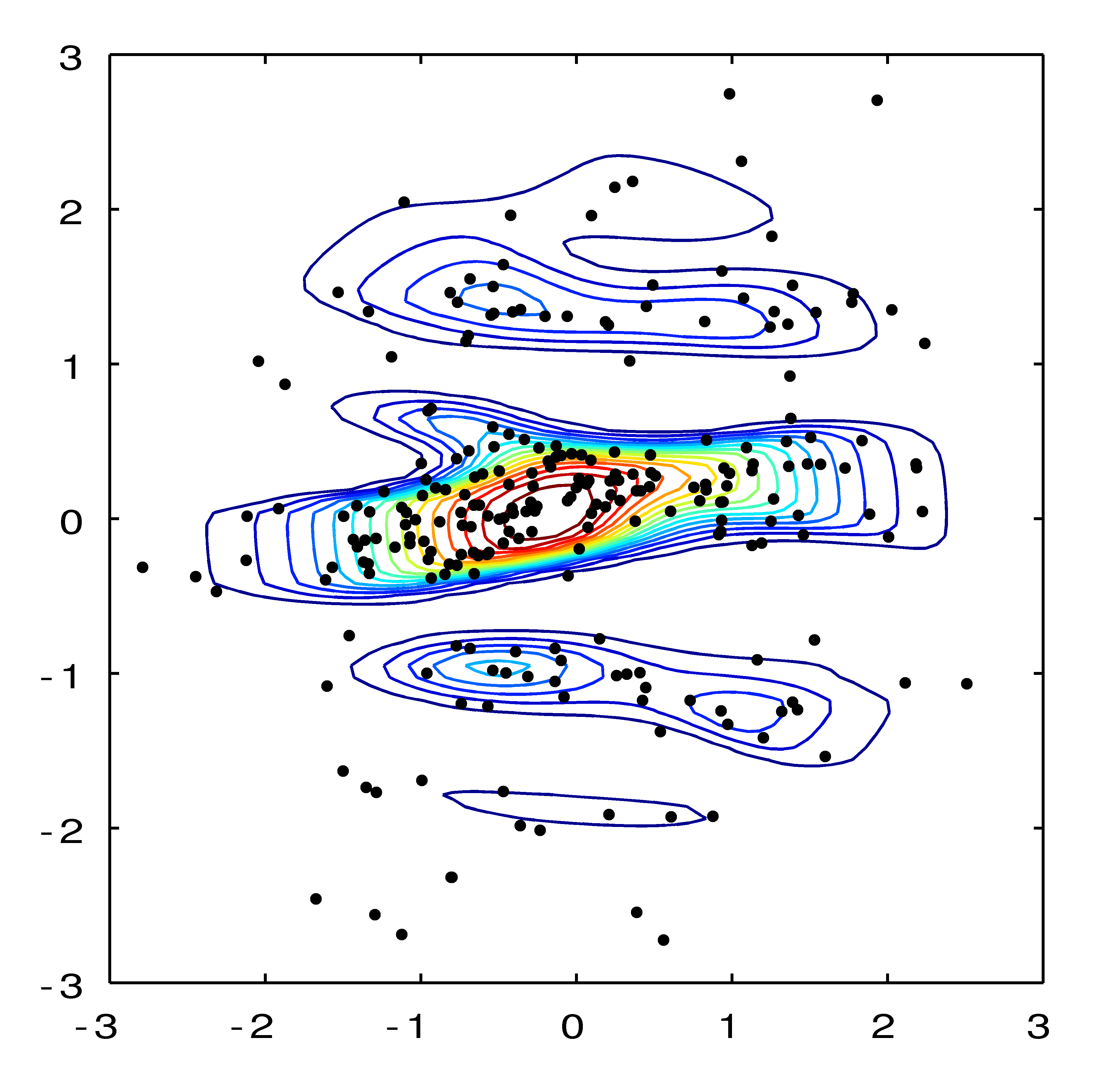}%
    }%
    \caption{Four samples from the GPDS prior are shown, with 250 data
      samples.  The contour lines show the approximate unnormalized
      densities.  In each case the base density is the zero-mean circular
      Gaussian with unit variance.  The mean function is set to zero.  The
      covariance function is the squared exponential:
      $C(\bx,\bx')=\alpha^2~\exp(-\half\sum_d (x_d-x_d')^2/\ell_d^2)$,
      with parameters varied as labeled in each subplot.  $\Phi(\cdot)$ is
      the logistic function in these plots.}
    \label{fig:prior-samples}
  \end{figure}

  We construct a map from the function~$g(\bx)$ to a probability density
  function\footnote{We will use the word ``density,'' according to the idea
    that~$\mcX$ is~$\reals^D$.  However, this construction would provide a
    distribution over probability mass functions for
    countable~$\mcX$.}~$f(\bx)$ via
  \begin{align}
    \label{eqn:transformation}
    f(\bx) &= \frac{1}{\mcZ_\pi[\bg]} \, \Phi(g(\bx)) \, \pi(\bx\given\psi)
  \end{align}
  where~$\pi(\bx\given\psi)$ is a parametric \textit{base density} that
  corresponds to an arbitrary base probability measure on~$\mcX$, with
  hyperparameters~$\psi$. The
  function~${\Phi(\cdot):\reals\rightarrow(0,1)}$ is a positive function
  with upper bound~$1$.  We use the bold notation~$\bg$ to refer to the
  function~$g(\bx)$ compactly as a vector of (infinite) length on which it is
  possible to perform inference.  The normalisation
  constant~$\mcZ_\pi[\bg]$ is a functional of~$g(\bx)$:
  \begin{align}
    \label{eqn:normalization-constant}
    \mcZ_\pi[\bg] &= \int\!\mathrm{d}x'\;\Phi(g(x')) \, \pi(x'\given\psi).
  \end{align}
  We include the subscript~$\pi$ to indicate implicit dependence on the
  density~$\pi(\bx)$.  Through the map defined by
  Equation~\ref{eqn:transformation}, a Gaussian process provides a prior
  distribution over normalised probability density functions
  on~$\mcX$. Figure~\ref{fig:prior-samples} shows several realisations of
  densities from this prior, along with sample data.

  Although we only require that the function~$\Phi(\cdot)$ be positive and
  bounded, it is convenient for inference if it is a bijective map
  between~$\reals$ and~$(0,1)$.  If~$\Phi(\cdot)$ is bijective then each
  function that maps~$\mcX$ to~$(0,1)$ corresponds to a unique
  realisation~$g(\bx)$ from the Gaussian process.  Sigmoids, such as the
  cumulative normal distribution function and the logistic function, are
  bijective functions with this domain and range.  We take~$\Phi(\cdot)$ to
  be the logistic function, i.e., $\Phi(z) = 1/(1+\exp(-z))$.

  \afterpage{\clearpage}
  
  \section{Generating data from the prior}
  \label{sec:prior-samples}

  We can use rejection sampling to simulate samples from a common density
  drawn from the the prior described in Section~\ref{sec:density-prior}.  A
  rejection sampler requires a proposal density that upper bounds the
  unnormalised density of interest.  In this case, the proposal density
  is~$\pi(\bx\given\psi)$ and the unnormalised density of interest
  is~$\Phi(g(\bx))\,\pi(\bx\given\psi)$.  We assume that it is possible to
  draw samples directly from~$\pi(\bx\given\psi)$.

\begin{figure}[t]
    \centering
    \graphicspath{{figures/prior-algorithm/}}%
    \subfloat[][{\scriptsize Initial state}]{%
      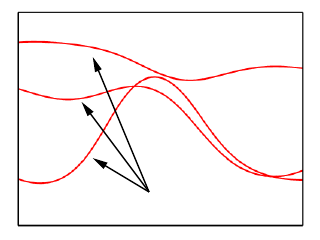%
    }%
    \subfloat[][{\scriptsize Draw $\btx_1 \sim \pi(\bx)$}]{%
      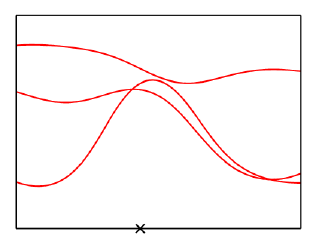%
    }%
    \subfloat[][{\scriptsize Sample $g_1$ from GP}]{%
      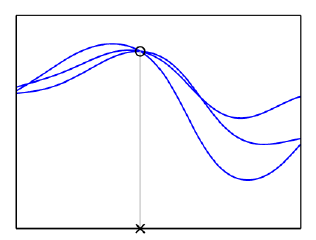%
    }%
    \subfloat[][{\scriptsize Draw {$u_1 \sim \distUni(0,1)$}}]{%
      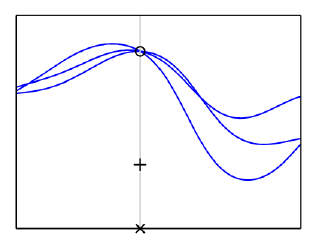%
    }\\%
    \subfloat[][{\scriptsize Iteration $r=2$}]{%
      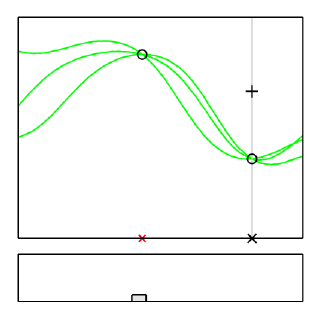%
    }%
    \subfloat[][{\scriptsize Iteration $r=3$}]{%
      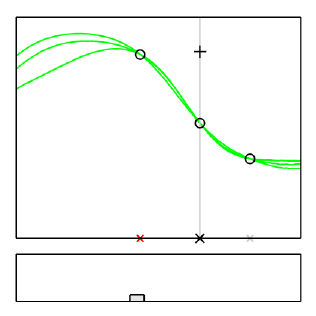%
    }%
    \subfloat[][{\scriptsize Iteration $r=4$}]{%
      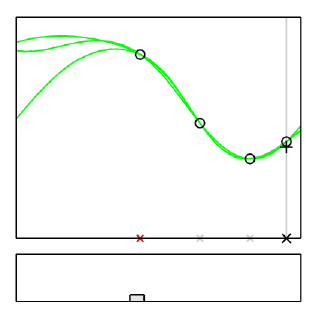%
    }%
    \subfloat[][{\scriptsize Iteration $r=100$}]{%
      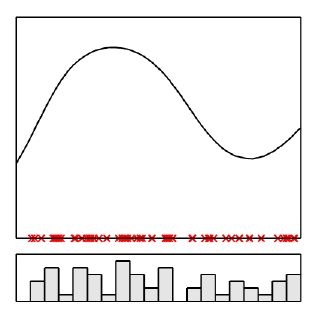%
    }%
    \label{fig:prior-algorithm}
  \end{figure}

  If~$g(\bx)$ were known, rejection sampling would proceed as follows: first
  generate proposals~$\{ \btx_r \}$ from the base
  density~$\pi(\bx\given\psi)$.  The proposal~$\btx_r$ would be accepted if a
  variate~$u_r$ drawn uniformly from~$(0,1)$ was less
  than~$\Phi(g(\btx_r))$.  These samples would be exact in the sense that
  they were not biased by the starting state of a finite Markov chain.
  However, in the GPDS,~$g(\bx)$ is not known: it is a random function drawn
  from a Gaussian process prior.  We can nevertheless use rejection
  sampling by ``discovering''~$g(\bx)$ as we proceed at just the places we
  need to know it, by sampling from the prior distribution of the latent
  function.  As the values of~$g(\bx)$ evaluated at the~$\{\btx_r\}$ are
  consistent with a single draw of the whole function,
  the samples are exact.  This type
  of retrospective sampling trick has been used in a variety of MCMC
  algorithms for infinite-dimensional models
  \cite{beskos-etal-2006a,papaspiliopoulos-roberts-2008a}.
  Figure~\ref{fig:prior-algorithm} shows the generative procedure
  graphically.
  
  In practice, we generate the samples sequentially, as in
  Algorithm~\ref{alg:prior-sample}, so that we may be assured of having as
  many accepted samples as we require.  In each loop, a proposal is drawn
  from the base density~$\pi(\bx\given\psi)$ and the function~$g(\bx)$ is
  sampled from the Gaussian process at this proposed coordinate,
  conditional on all the function values already sampled.  We will call
  these data the \textit{conditioning set} for the function~$g(\bx)$ and will
  denote the conditioning inputs as~$\biX$ and the conditioning function
  values as~$\biG$.  After the function is sampled, a variate is drawn
  uniformly from~$(0,1)$ and compared to the~$\Phi$-squashed function at
  the proposal location.  If the uniform variate falls below $\Phi(g(\bx))$
  then we accept the proposal, otherwise we reject.  The proposals and
  their function values are added into the conditioning set regardless of
  whether that proposal was accepted or rejected.  The loop repeats until
  we have as many acceptances as are required.

  The sequential procedure is infinitely exchangeable; the probability of
  the data is the same under reordering.  First, the base density draws are
  i.i.d..  Second, conditioned on the proposals from the base density, the
  Gaussian process is a simple multivariate Gaussian distribution, which is
  exchangeable in its components.  Finally, conditioned on the draw from
  the Gaussian process, the acceptance/rejection steps are independent
  Bernoulli samples, and the overall procedure is exchangeable.  This
  property ensures that the sequential procedure generates data from the
  same distribution as the simultaneous procedure described above.  More
  broadly, exchangeable priors are useful in Bayesian modeling because we
  may consider the data conditionally independent, given the latent
  density.

  \begin{algorithm}[t] 
    \begin{algorithmic}[1]

  \Ensure%
  \begin{minipage}[t]{\linewidth}
    \begin{itemize*}
    \item Number of samples to draw~$N$
    \item Gaussian process covariance function~$C(\bx,\bx'\,;\,\theta)$
    \item Base density~$\pi(\bx \given \psi)$
    \end{itemize*}
  \end{minipage}

  \Require%
  \begin{minipage}[t]{\linewidth}
    \begin{itemize}
    \item $N$ samples $\mcD = \{\bx_n\}^N_{n=1}$ from a random density
      drawn from the prior.
    \end{itemize}
  \end{minipage}
  \vskip 0.25cm

  \State $\biX \gets \emptyset$, $\biG \gets \emptyset$
  \Comment{Initially the conditioning sets are empty.}

  \State $\mcD \gets \emptyset$
  \Comment{Initialize the set to be returned.}

  \State $r \gets 0$
  \Comment{Count the number of proposals.}
  
  \Repeat
  
  \State $\btx_r \sim \pi(\bx \given \psi)$
  \Comment{Draw a proposal.}

  \State $g(\btx_r) \sim \distGP(g \given \btx_r, \biX, \biG, \theta)$
  \Comment{Sample from the GP at the proposal.}

  \State $u_r \sim \distUni(0,1)$
  \Comment{Draw uniformly on~$(0,1)$.}
  
  \If{$u_r < \Phi(g(\btx_r))$}
  \Comment{Rejection sampling acceptance rule.}

  \State $\mcD \gets \mcD \cup \btx_r$
  \Comment{Store the proposal.}

  \EndIf

  \State $\biX \gets \biX \cup \btx_r$, $\biG \gets \biG \cup g(\btx_r)$
  \Comment{Update the conditioning sets, even on rejections.} 

  \State $r \gets r + 1$
  
  \Until{$||\mcD|| = N$} \Comment{Loop until $N$ samples are accepted.}

  \State \Return $\mcD$
  
\end{algorithmic}

    \caption{Generate $N$ exact samples from a density drawn from the
      prior}
    \label{alg:prior-sample}
  \end{algorithm}
  
  \section{Inference}
  We now consider the problem of inference with the GPDS.  We observe~$N$
  data~$\mcD=\{\bx_n\}^N_{n=1}$ that we model as having been drawn
  independently from an unknown density~$f(\bx)$.  We place the GPDS prior of
  Section~\ref{sec:density-prior} on $f(\bx)$.  The posterior on~$\bg$ is
  given by Bayes' theorem:
  \begin{align}
    \label{eqn:posterior}
    p(\bg\given\mcD,\theta) &=
    \frac{
      p(\bg\given\theta) \; 
      (\mcZ_\pi[\bg])^{-N}
      \prod_{n=1}^{N}\Phi(g(\bx_n)) \; \pi(\bx_n \given \psi)
    }{
      \int \mathrm{d}\bg' \;
      p(\bg'\given\theta) \;
      (\mcZ_\pi[\bg'])^{-N}
      \prod_{n=1}^{N}
      \Phi(g'(\bx_n)) \; \pi(\bx_n \given \psi)}.
  \end{align}
  Even with Markov chain Monte Carlo, inference in this model is difficult.
  Evaluating the posterior requires computing two difficult integrals, the
  denominator and the normalisation constant~$\mcZ_\pi[\bg]$.  It is common
  for the marginal likelihood in the denominator of the posterior to be
  intractable; MCMC methods such as Metropolis--Hastings are well-suited
  for this situation. Posteriors such as Equation~\ref{eqn:posterior} with
  difficult sums in both the numerator and denominator are called
  \textit{doubly-intractable}. Doubly-intractable posterior distributions
  appear most frequently when performing inference in undirected graphical
  models, where the partition function can be difficult to
  evaluate~\citep{moller-etal-2006a,murray-etal-2006a}.

  To see the difficulty concretely, consider a na\"{i}ve
  Metropolis--Hastings Markov chain on $\bg$, with proposal density
  $q(\bhg\leftarrow\bg)$:
  \begin{align}
    a_{\textsf{na{\"i}ve}} &=
    \frac{
      q(\bg\leftarrow\bhg) \;
      p(\bhg\given\theta)
      }{
      q(\bhg\leftarrow\bg) \;
      p(\bg\given\theta)
      }
    \left(\frac{\mcZ_\pi[\bhg]}{\mcZ_\pi[\bg]}\right)^{N}
    \prod^N_{n=1}
    \frac{\Phi(\hg(\bx_n)) \; \pi(\bx_n \given \psi)}
    {\Phi(g(\bx_n)) \; \pi(\bx_n \given \psi)}.
  \end{align}
  The functions $\bg$ and $\bhg$ are infinite-dimensional objects, which
  cannot be evaluated everywhere in practice. In other contexts, such as
  Gaussian process classification, it is possible to construct the proposal
  density such that the acceptance ratio only depends on the functions at
  $\{\bx_n\}$. Here, the intractable ratio of normalising constants makes it
  impossible to evaluate the acceptance ratio without knowing the $\bg$ and
  $\bhg$ everywhere. We present two Markov chain Monte Carlo algorithms
  that sidestep this difficulty.  The equilibrium distribution in both
  cases is the posterior in Equation~\ref{eqn:posterior}, and both
  algorithms take advantage of the exact data generation procedure
  described in Section~\ref{sec:prior-samples}.
    
  \subsection{Exchange sampling}
  \label{sec:exchange-sampling}
  Exchange sampling \citep{murray-etal-2006a,murray-2007a} is a variant of
  the Metropolis--Hastings method that enables sampling from
  doubly-intractable posterior distributions, subject to the requirement
  that exact samples can be generated from the model.  The procedure is
  a simpler alternative to the auxiliary variable method of
  \citet{moller-etal-2006a}.  Exchange sampling introduces additional state
  into the Markov chain that is chosen so that the intractable constants
  cancel out of the Metropolis--Hastings acceptance ratio.
  \citet{murray-etal-2006a} used exchange sampling to infer the coupling
  parameters of Ising models where exact data could be generated
  via coupling from the past~\citep{propp-wilson-1996a}.  In the GPDS we
  generate exact samples via the rejection method of
  Section~\ref{sec:prior-samples}.

  \begin{algorithm}[t] 
    \begin{algorithmic}[1]

  \Ensure%
  \begin{minipage}[t]{\linewidth}
    \begin{itemize*}
    \item Number of MCMC iterations $R$
    \item Observed data $\mcD = \{\bx_n\}^N_{n=1}$
    \item Gaussian process covariance function~$C(\bx,\bx'\,;\,\theta)$
    \item Base density $\pi(\bx \given \psi)$
    \end{itemize*}
  \end{minipage}

  \Require%
  \begin{minipage}[t]{\linewidth}
    \begin{itemize}
    \item $R$ conditioning sets of function inputs and
      outputs~$\{\biX^{(r)}, \biG^{(r)}\}^R_{r=1}$
    \end{itemize}
  \end{minipage}
  \vskip 0.25cm
  \State $\{g(\bx_n)\}^N_{n=1} \sim
  \distGP(g\given\mcD,\theta)$
  \Comment{Initialise the function at the data.}

  \State $\biX^{(1)}\gets\{\bx_n\}^N_{n=1}$,
  $\biG^{(1)}\gets\{g(\bx_n)\}^N_{n=1}$
  \Comment{Initialise conditioning sets.}

  \For{$r\gets 1\ldots R$}
  \Comment{Take $R$ exchange sampling steps.}
  
  \State $\{\hg(\bx_n)\}^N_{n=1} \sim \distGP(g\given\mcD,\theta)$
  \Comment{Draw a new function at the data.}

  \State $\bhiX\gets\{\bx_n\}^N_{n=1}$,
  $\bhiG\gets\{\hg(\bx_n)\}^N_{n=1}$
  \Comment{Initialise proposal conditioning sets.}
  \label{algstep:fantasy-start}

  \State $\mcW\gets\emptyset$
  \Comment{Initialise empty fantasy data set.}

  \Repeat
  \Comment{Run the rejection sampling loop.}

  \State $\btw \sim \pi(\bx\given\psi)$
  \Comment{Draw a proposal from the base density.}

  \State $\hg(\btw) \sim \distGP(\hg\given\btw,\bhiX,\bhiG,\theta)$
  \Comment{Draw the function value at the proposal.}

  \State $u_{\sf{fant}} \sim \distUni(0,1)$
  \Comment{Draw a uniform random variate on $(0,1)$.}
  \If{$u_{\sf{fant}} < \Phi(\hg(\btw))$}
  \Comment{Rejection sampling acceptance rule.}
  \label{algstep:fantasy-acceptance}

  \State $\mcW \gets \mcW \cup \btw$
  \Comment{Keep the fantasy.}

  \EndIf

  \State $\bhiX\gets\bhiX\cup\btw$, $\bhiG\gets\bhiG\cup\hg(\btw)$
  \Comment{Add proposals to the conditioning sets.}
  \label{algstep:fantasy-store}

  \Until{$|\mcW| = N$}
  \Comment{Loop until $N$ fantasies are accepted.}
  \label{algstep:fantasy-end}
  
  \State $\{g(\bw_n)\}^N_{n=1} \sim \distGP(g\given\mcW, \biX^{(r)},
  \biG^{(r)})$
  \Comment{Sample the current func. at the fantasies.}
  \label{algstep:current-at-fantasy}

  \State $a_{\sf{exch}} \gets \displaystyle \prod^N_{n=1}
  \frac{\Phi(\hg(\bx_n))\;\Phi(g(\bw_n))}
  {\Phi(g(\bx_n))\;\Phi(\hg(\bw_n))}$
  \Comment{Calculate the acceptance ratio.}
  
  \State $u_{\sf{mh}} \sim \distUni(0,1)$
  \Comment{Draw a uniform random variate on $(0,1)$.}

  \If{$u_{\sf{mh}} < a_{\sf{exch}}$}
  \Comment{Apply the Metropolis--Hastings acceptance rule.}
  \label{algstep:mh-acceptance}

  \State $\biX^{(r+1)}\gets\bhiX$, $\biG^{(r+1)}\gets\bhiG$
  \Comment{Keep the new function data.}

  \Else
  \State $\biX^{(r+1)}\gets\biX^{(r)}\cup\{\bw_n\}^N_{n=1}$
  \Comment{Add the fantasy evaluations to the current state.}
  \label{algstep:current-store}

  \State $\biG^{(r+1)}\gets\biG^{(r)}\cup\{g(\bw_n)\}^N_{n=1}$

  \EndIf

  \EndFor
  \State \Return $\{\biX^{(r)}, \biG^{(r)}\}^R_{r=1}$

\end{algorithmic}

    \caption{Simulate $R$ steps of an exchange sampling Markov chain on $p(\bg\given\mcD)$}
    \label{alg:exchange-sampling}
  \end{algorithm}
  
  Initially, we apply exchange sampling to the posterior on~$\bg$ using the
  Gaussian process prior as the proposal distribution, i.e.,~$q(\bhg\leftarrow\bg) = p(\bhg\given\theta)$.  The joint distribution
  over the data $\mcD$, the current Markov state $\bg$ and the proposal
  $\bhg$ is augmented with~$N$ ``fantasy data''~$\mcW = \{\bw_n\}^N_{n=1}$.
  These fantasy data live on the same space~$\mcX$ as the true data, but
  are drawn from the distribution implied by the proposal~$\bhg$.  The
  augmented joint distribution is
  \begin{align}
    \label{eqn:simple-exchange-joint}
    p(\bg,\mcD, \bhg, \mcW\given\theta, \psi)
    &=
    p(\bg\given\theta)
    p(\{\bx_n\}^N_{n=1}\given\bg,\psi)
    p(\bhg\given\theta) p(\{\bw_n\}^N_{n=1}\given\bhg,\psi).
  \end{align}

  Given the current state~$\bg$, we jointly propose $\bhg$ and $\mcW$ by
  using Algorithm~\ref{alg:prior-sample}.  This algorithm
  simultaneously draws $\bhg$ from the prior and generates the $N$ fantasy
  data $\mcW$.  We then propose swapping~$\bg$ with~$\bhg$.  The acceptance
  ratio of the swap proposal is the ratio of the joint density in
  Equation~\ref{eqn:simple-exchange-joint} under each setting:
  \begin{align}
    a_{\textsf{exch}} &= \frac{
      \cancel{p(\bhg\given\theta)} \; p(\{\bx_n\}^N_{n=1}\given\bhg,\psi) \;
      \cancel{p(\bg\given\theta)} \; p(\{\bw_n\}^N_{n=1}\given\bg,\psi)
    }{
      \cancel{p(\bg\given\theta)} \; p(\{\bx_n\}^N_{n=1}\given\bg,\psi) \;
      \cancel{p(\bhg\given\theta)} \; p(\{\bw_n\}^N_{n=1}\given\bhg,\psi)
    }\notag\\
    \label{eqn:independence-acceptance-ratio}
    &= \frac{
      \cancel{\mcZ_\pi[\bg]^{N}}
      \cancel{\mcZ_\pi[\bhg]^{N}}
      \prod^N_{n=1}\Phi(\hg(\bx_n)) \; \cancel{\pi(\bx_n\given\psi)}
      \prod^N_{n=1}\Phi(g(\bw_n)) \; \cancel{\pi(\bw_n\given\psi)}
      }{
      \cancel{\mcZ_\pi[\bg]^{N}} 
      \cancel{\mcZ_\pi[\bg']^{N}}
      \prod^N_{n=1}\Phi(g(\bx_n)) \; \cancel{\pi(\bx_n\given\psi)}
      \prod^N_{n=1}\Phi(\hg(\bw_n)) \; \cancel{\pi(\bw_n\given\psi)}
      }\notag\\
    &= \prod^N_{n=1}\frac{\Phi(\hg(\bx_n)) \; \Phi(g(\bw_n))}
    {\Phi(g(\bx_n)) \; \Phi(\hg(\bw_n))}.
  \end{align}
  The normalisation constants cancel out, and the functions $g(\bx)$ and
  $\hg(\bx)$ need only be sampled from the Gaussian process at a finite number
  of locations.

\begin{figure}[t!]
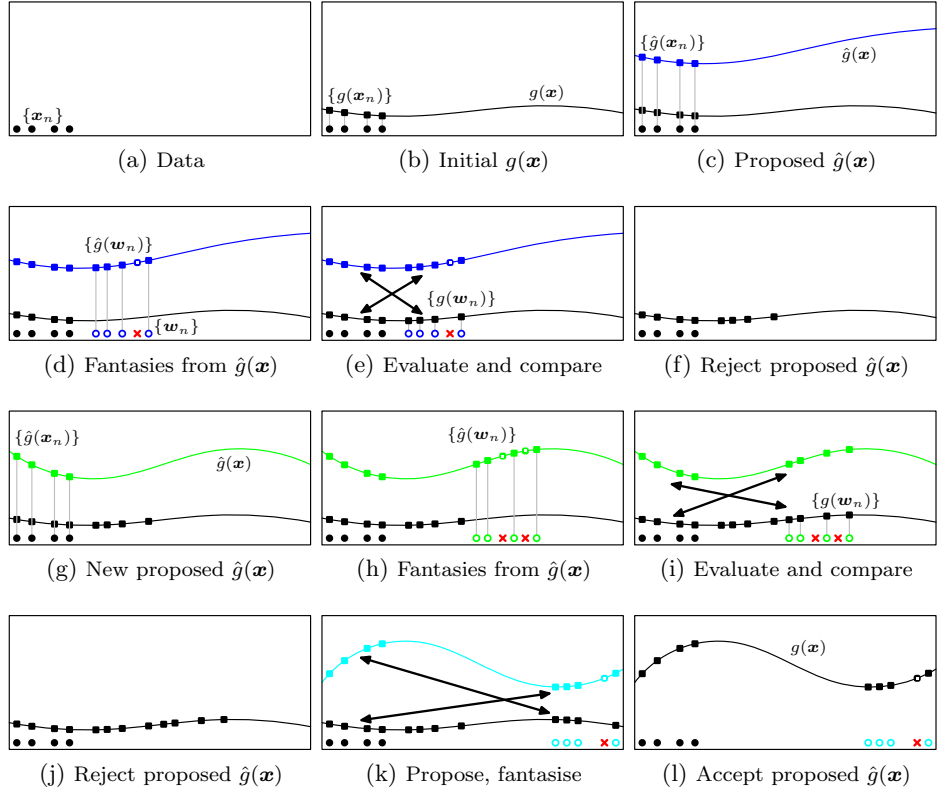

    \centering%
    \subfloat[][{\scriptsize Data}]{%
      \includegraphics[width=0.32\textwidth]{figures/exchange/exchange-mh.1}%
    }~%
    \subfloat[][{\scriptsize Initial $g(\bx)$}]{%
      \includegraphics[width=0.32\textwidth]{figures/exchange/exchange-mh.2}%
    }~%
    \subfloat[][{\scriptsize Proposed $\hg(\bx)$}]{%
      \includegraphics[width=0.32\textwidth]{figures/exchange/exchange-mh.3}%
    }\\
    \subfloat[][{\scriptsize Fantasies from $\hg(\bx)$}]{%
      \includegraphics[width=0.32\textwidth]{figures/exchange/exchange-mh.4}%
    }~%
    \subfloat[][{\scriptsize Evaluate and compare}]{%
      \includegraphics[width=0.32\textwidth]{figures/exchange/exchange-mh.5}%
    }~%
    \subfloat[][{\scriptsize Reject proposed $\hg(\bx)$}]{%
      \includegraphics[width=0.32\textwidth]{figures/exchange/exchange-mh.6}%
    }\\
    \subfloat[][{\scriptsize New proposed $\hg(\bx)$}]{%
      \includegraphics[width=0.32\textwidth]{figures/exchange/exchange-mh.7}%
    }~%
    \subfloat[][{\scriptsize Fantasies from $\hg(\bx)$}]{%
      \includegraphics[width=0.32\textwidth]{figures/exchange/exchange-mh.8}%
    }~%
    \subfloat[][{\scriptsize Evaluate and compare}]{%
      \includegraphics[width=0.32\textwidth]{figures/exchange/exchange-mh.9}%
    }\\
    \subfloat[][{\scriptsize Reject proposed $\hg(\bx)$}]{%
      \includegraphics[width=0.32\textwidth]{figures/exchange/exchange-mh.10}%
    }~%
    \subfloat[][{\scriptsize Propose, fantasise }]{%
      \includegraphics[width=0.32\textwidth]{figures/exchange/exchange-mh.11}%
    }~%
    \subfloat[][{\scriptsize Accept proposed $\hg(\bx)$}]{%
      \includegraphics[width=0.32\textwidth]{figures/exchange/exchange-mh.12}%
    }%
    \caption[Illustration of exchange sampling for the GPDS]{A cartoon of
      three exchange sampling transitions.  In the first two transitions,
      the proposals are rejected to demonstrate the expanding Markov state
      due to retrospective sampling of~$g(\bx)$. The third proposal is
      accepted and the previously-accumulated state is discarded. (a)~The
      observed data, illustrated as~\protect
      \raisebox{0.5ex}{\includegraphics{figures/exchange/exchange-mh.13}}\,\protect.  (b)~The
      initial~$g(\bx)$ evaluated at the data, shown as~\protect
      \raisebox{0.5ex}{\includegraphics{figures/exchange/exchange-mh.14}}\,\protect. (c)~The
      proposed~$\hg(\bx)$ evaluated at the data, shown as~\protect
      \raisebox{0.5ex}{\includegraphics{figures/exchange/exchange-mh.18}}\,\protect.
      (d)~Fantasies are drawn from~$\hg(\bx)$, illustrated as~\protect
      \raisebox{0.5ex}{\includegraphics{figures/exchange/exchange-mh.16}}\,\protect.  There is
      one rejected proposal, shown as~\protect
      \raisebox{0.5ex}{\includegraphics{figures/exchange/exchange-mh.17}}\,\protect, with the
      corresponding function value illustrated as~\protect
      \raisebox{0.5ex}{\includegraphics{figures/exchange/exchange-mh.15}}\,\protect.
      (e)~$g(\bx)$ is evaluated at the fantasies and the two explanations
      are compared using
      Equation~\ref{eqn:independence-acceptance-ratio}.  (f)~The
      proposal~$\hg(\bx)$ is rejected.  The Markov state expands to include
      the fantasies.  (g)~A new~$\hg(\bx)$, shown in green, is evaluated at
      the data.  (h)~Fantasies are drawn from~$\hg(\bx)$ (two rejected
      fantasy proposals).  (i)~$g(\bx)$ is evaluated at the fantasies and
      the explanations are compared.  (j)~The proposal was rejected.  The
      Markov state expands to twelve function evaluations.  (k)~Skipping
      the intermediate steps, propose, fantasise, evaluate and compare,
      using the function shown in cyan.  (l)~The proposal is accepted and
      made the new~$g(\bx)$.  All of the information about the old function
      is thrown away, but the new~$g(\bx)$ must keep information in its
      conditioning set about the fantasies it generated.}%
    \label{fig:exchange-cartoon}%
  \end{figure}

  Algorithm~\ref{alg:exchange-sampling} shows the exchange sampling
  inference procedure for the GPDS.  Some amount of bookkeeping is required
  for this procedure to be valid.  Specifically, once something is learned
  about a particular function~$g(\bx)$, i.e., sampled from the Gaussian
  process, it cannot be forgotten until that~$g(\bx)$ is discarded.  For
  example, when fantasy data is generated from~$\hg(\bx)$, as in
  steps~\ref{algstep:fantasy-start}~to~\ref{algstep:fantasy-end}, even
  if~$\btx$ is rejected in step~\ref{algstep:fantasy-acceptance},
  the~$(\btx, g(\btx))$ pair must be stored in the conditioning set
  (step~\ref{algstep:fantasy-store}).  If the proposed~$\hg(\bx)$ is
  ultimately rejected by step~\ref{algstep:mh-acceptance}, only then can
  the conditioning set for~$\hg(\bx)$ be discarded.  Similarly, when the
  current Markov state~$g(\bx)$ is sampled from the Gaussian process at the
  fantasy data in step~\ref{algstep:current-at-fantasy}, this information
  must be kept if the proposal is rejected
  (step~\ref{algstep:current-store}).  Thus
  step~\ref{algstep:current-store} expands the Markov state with every
  rejection, as information about the current~$g(\bx)$ accumulates.  When
  the proposal~$\hg(\bx)$ is accepted, the Markov state reduces in size, as
  fewer points will typically have been sampled from~$\hg(\bx)$. An example
  sequence of rejections and an acceptance is illustrated in
  Figure~\ref{fig:exchange-cartoon}.

  \afterpage{\clearpage}

  \subsubsection{Improving the acceptance rate}
  \label{sec:exchange-control}
  For clarity, we introduced the algorithm with~${q(\bhg\leftarrow\bg) =
    p(\bhg\given\theta)}$, but this proposal is a poor choice in practice.
  To achieve a better acceptance rate, it is better to make conservative,
  perturbative proposals.  This can be achieved by introducing a set of~$B$
  ``control points'' in~$\mcX$, denoted~$\mcC = \{\bx_b\in\mcX\}^B_{b=1}$.
  These control points have associated function values, which we denote
  as~${\mcG=\{g(\bx_b)\}^B_{b=1}}$.  We assume that~$\mcC$ is a superset of
  the observed data, i.e.,~$\mcD\subseteq\mcC$.  The function values at the
  control points are explicitly included in the Markov state and all
  retrospective function draws condition on these points.  New discoveries
  about the function continue to accumulate in the conditioning sets as
  before.  The difference now is that the conditioning sets are initialised
  with the control points and small, perturbative proposals can be made on
  the function values at those initial points.
  
  To make this construction explicit,
  Equation~\ref{eqn:simple-exchange-joint} is extended to
  \begin{multline}
    \label{eqn:control-joint}
    p(\mcG,\bg_{\backslash\mcC}, \mcD, \hmcG,
    \bhg_{\backslash\mcC}, \mcW\given\mcC, \theta,\psi) = \\
    \distGP(\mcG\given\mcC,\theta)\;
    \distGP(\bg_{\backslash\mcC}\given\mcG,\theta)\,
    \mcZ_{\pi}[\bg]^{-N}
    \left[\prod^N_{n=1}\Phi(g(\bx_n))\;\pi(\bx_n\given\psi)\right] \\
    \times q(\hmcG\leftarrow\mcG)\;
    \distGP(\bhg_{\backslash\mcC}\given\hmcG,\theta)\;
    \mcZ_{\pi}[\bhg]^{-N}
    \prod^N_{n=1}\Phi(\hg(\bw_n))\;\pi(\bw_n\given\psi),
  \end{multline}
  where~$\hmcG$ indicates the proposal of the function values at the
  control points, i.e.~${\hmcG=\{\hg(\bx_b)\}^B_{b=1}}$, and
  $\bg_{\backslash \mcC}$ denotes the function values, excluding those
  at~$\mcC$.  The proposal density~$q(\hmcG\leftarrow\mcG)$ can be chosen
  to take smaller steps than the prior draws of the previous section.  With
  the joint distribution in Equation~\ref{eqn:control-joint}, and using the
  conditional retrospective exchange sampling as before, the acceptance
  ratio of exchanging the pair~$(\mcG,\bg_{\backslash\mcC})$
  for~$(\hmcG,\bhg_{\backslash\mcC})$ is
  \begin{align}
    a_{\textsf{exch-cp}} &=
    \frac{
      q(\mcG\leftarrow\hmcG)\;
      \distGP(\hmcG\given\mcC,\theta)
    }{
      q(\hmcG\leftarrow\mcG)\;
      \distGP(\mcG\given\mcC,\theta)
    } 
    \prod^N_{n=1}\frac{
      \Phi(\hg(\bx_n))\;
      \Phi(g(\bw_n))
    }{
      \Phi(g(\bx_n))\;
      \Phi(\hg(\bw_n))
    }.
  \end{align}
  Superficially, this might seem similar to the knot-based imputation
  method of \citet{tokdar-2007a}.  However, whereas \citet{tokdar-2007a}
  uses knots as a finite-dimensional approximation, we use the control
  points simply to constrain the proposal distribution.  The control points
  only initialise the retrospective sampling procedure.  As we enforce a
  Gaussian process prior on the function values of the control points, the
  inference procedure still yields the correct posterior distribution on
  the uncompromised fully-nonparametric Gaussian process density sampler
  model.  The number and locations of the control points are free
  parameters.
  
  A new function is proposed by first choosing values at the control points
  close to the existing function.  The remainder of the function is drawn
  from the prior, conditioned on the values at the control points. We
  always include the locations of the observed data as control points,
  i.e.,~$\mcD \subset \mcC$.  This is not required for the algorithm to be
  valid, but is convenient as all proposed functions must be evaluated at
  the data in any case. Taking account of the arbitrary proposal density at
  the control points, $q(\{\hg(\bx_k)\}^K_{k=1} \leftarrow
  \{g(\bx_k)\}^K_{k=1})$, the exchange sampling acceptance ratio becomes
  \begin{multline}
    a_{\textsf{exch-control}}
    = \frac{
      q(\{g(\bx_k)\}^K_{k=1}\leftarrow\{\hg(\bx_k)\}^K_{k=1}) \;
      p(\{\hg(\bx_k)\}^K_{k=1}\given\theta)
      }{
      q(\{\hg(\bx_k)\}^K_{k=1}\leftarrow\{g(\bx_k)\}^K_{k=1}) \;
      p(\{g(\bx_k)\}^K_{k=1}\given\theta)
    }\\\times
    \prod^N_{n=1}\frac{\Phi(\hg(\bx_n)) \; \Phi(g(\bw_n))}
	 {\Phi(g(\bx_n)) \; \Phi(\hg(\bw_n))}.
  \end{multline}
  The functions drawn from the Gaussian process must still be evaluated at
  a larger conditioning set that includes the locations of fantasies. As
  before, these can be drawn ``retrospectively'' as needed, but now these
  Gaussian process samples are conditioned on the values at the control
  points.
  
  \subsubsection{Hyperparameter inference}
  \label{sec:exchange-hyper}
  One of the benefits of the Bayesian approach is the ability to perform
  hierarchical inference.  In this case, it allows us to infer the
  hyperparameters $\theta$ of the Gaussian process and the hyperparameters
  $\psi$ of the base density.  We augment the exchange sampling algorithm
  slightly to sample from the posterior on hyperparameters: before
  proposing a new function $\hg(\bx)$, we propose new hyperparameters
  $\htheta$ and $\hpsi$ from a proposal density $q(\htheta,\hpsi \leftarrow
  \theta,\psi)$.  When samples of the new function are drawn, it is done
  with these proposed hyperparameters.  The new joint distribution is
  \begin{multline}
    p(\bg,\{\bx_n\}^N_{n=1}, \theta, \psi, \bhg, \{\bw_n\}^N_{n=1}, \htheta, \hpsi)
    = \\
    p(\theta,\psi) \; p(\bg\given\theta) \;
    p(\{\bx_n\}^N_{n=1}\given\bg,\psi) \\
    \times q(\htheta,\hpsi\leftarrow\theta,\psi) \;
    p(\bhg\given\htheta) \;
    p(\{\bw_n\}^N_{n=1}\given\bhg,\hpsi)
  \end{multline}
  where~$p(\theta,\psi)$ is an appropriate hyperprior.  The proposal is now
  to exchange the triplets~$(\bg,\theta,\psi)$ and~$(\bhg,\htheta,\hpsi)$.
  The acceptance of this swap has Metropolis--Hastings ratio
  \begin{multline}
    a_{\textsf{exch-hyper}}
    = \frac{
      q(\theta,\psi\leftarrow\htheta, \hpsi) \;
      p(\htheta,\hpsi)
    }{
      q(\htheta,\hpsi\leftarrow\theta,\psi) \;
      p(\theta, \psi)
      }\\\times
    \prod^N_{n=1}
    \frac{
      \Phi(\hg(\bx_n)) \; \pi(\bx_n\given\hpsi) \;
      \Phi(g(\bw_n)) \; \pi(\bw_n\given\psi)
      }{
      \Phi(g(\bx_n))\; \pi(\bx_n\given\psi) \;
      \Phi(\hg(\bw_n)) \; \pi(\bw_n\given\hpsi)
      }.
  \end{multline}
  This acceptance ratio generalises straightforwardly to the case with
  control points discussed in Section~\ref{sec:exchange-control}.
  
  \subsubsection{Sampling from the predictive distribution}
  \label{sec:exchange-predictive}
  An important task for density inference is estimation of the predictive
  density.  The predictive distribution arises on data space when the
  posterior is integrated out.  For the GPDS, this density is
  \begin{align}
    \label{eqn:predictive-integral}
    p(\bx\given\mcD)
    &=
    \int\mathrm{d}\theta
    \int\mathrm{d}\psi
    \int\mathrm{d}\bg \;
    p(\bx\given\bg,\theta,\psi) \;
    p(\bg,\theta,\psi\given\mcD).
  \end{align}
  The predictive distribution can also be thought of as the distribution on
  the next datum to arrive, given the~$N$ already seen and taking
  uncertainty into account.  In the GPDS, the predictive density in
  Equation~\ref{eqn:predictive-integral} is not available analytically.  We
  nevertheless have all the tools in place to generate samples from the
  predictive distribution.  We do this by using the generative procedure of
  Section~\ref{sec:prior-samples} to generate additional data after each
  Metropolis--Hastings step.  We use a very similar method to
  Algorithm~\ref{alg:prior-sample}, but initialise the conditioning set
  using the current state of the Markov chain. 
  
  \subsection{Sampling over latent histories}
  An alternative to inference via exchange sampling is to model the
  \textit{latent history} of the generative process.  By using the GPDS
  prior to model the data, we are asserting that the data can be explained
  as the result of Algorithm~\ref{alg:prior-sample}.  However, we did not
  observe any of the intermediate states of the rejection sampling
  algorithm, such as the number and locations of the rejected proposals,
  and the value of the function sampled from the Gaussian process prior.
  Nevertheless, Algorithm~\ref{alg:prior-sample} provides a well-defined
  probabilistic model over both the observed data and this latent state.
  By modeling this larger joint distribution we can avoid evaluating the
  intractable normalisation constant that would otherwise appear in the
  likelihood function.
  
  We model the data ${\mcD=\{\bx_n\}^N_{n=1}}$ as having been generated
  exactly as in Algorithm~\ref{alg:prior-sample}, i.e., run until
  exactly~$N$ proposals were accepted.  The state space of the Markov chain
  on latent histories in the GPDS consists of: 1)~the values of the latent
  function~$g(\bx)$ at the data, denoted~${\mcG_N=\{g(\bx_n)\}^N_{n=1}}$,
  2)~the number of rejections~$M$, 3)~the locations of the $M$ rejected
  proposals, denoted~${\mcM=\{\bx_m\}^M_{m=1}}$, and 4)~the values of the
  latent function~$g(\bx)$ at the~$M$ rejected proposals, denoted~${\mcG_M
    =\{g(\bx_m)\}^M_{m=1}}$.  The joint distribution over the data and the
  ordered history of the GPDS generative procedure, given the
  hyperparameters, is
  \begin{multline}
    \label{eqn:gpds-lh-joint}
    p(\mcD, \mcG_N, \mcM,\mcG_M\given\theta,\psi) =
    \distGP(\mcG_N,\mcG_M\given\mcD, \mcM, \theta)\\
    \times \left[\prod^N_{n=1}\Phi(g(\bx_n))\,\pi(\bx_n\given\psi)\right]
    \prod^M_{n=1}(1-\Phi(g(\bx_m)))\,\pi(\bx_m\given\psi).
  \end{multline}

  We sample from the posterior distribution over all unknowns, which is
  proportional to the joint distribution with the observations, $\mcD$,
  clamped. The Markov chain algorithm applies three types of update in
  sequence:
  1)~modification of the number of rejections~$M$,
  2)~updating of the rejection locations~$\mcM$, and 3)~modification of the
  latent function values~$\mcG_M$ and~$\mcG_N$.  We will maintain an
  explicit ordering of the latent rejections for reasons of clarity,
  although this is not necessary due to exchangeability.  At any time we
  could propose a reshuffling of the latent history, subject to it ending
  in an acceptance, and this proposal would always be accepted, as the two
  permutations have the same probability under the model.

  Inference by Markov chain Monte Carlo of the history of a probabilistic
  computational procedure has been studied previously.
  \citet{beskos-etal-2006a} sampled from the state of a rejection sampler
  for diffusions.  \citet{murray-2007a}, who coined the phrase ``latent
  history,'' modeled data as having been the result of a Markov chain which
  had provably mixed via coupling from the past \citep{propp-wilson-1996a}.
  Another example is \citet{huber-wolpert-2009a}, who model the history of
  the Mat{\'e}rn Type III process to perform tractable inference.  The
  Church programming language \citep{goodman-etal-2008a} also exploits this
  idea, by treating probabilistic procedures as first class objects on
  which inference can be performed.

  \subsubsection{Modifying the number of latent rejections}
  We propose a new number of latent rejections~$\hM$ by drawing it from a
  proposal density $q(\hM~\leftarrow~M)$.  If~$\hM$ is greater
  than~$M$, we must also propose new rejections to add to the latent state.
  We take advantage of the exchangeability of the process to generate the
  new rejections: we imagine these proposals were made \textit{after} the
  last observed datum was accepted, and our proposal is to call them
  rejections and move them \textit{before} the last datum.  If~$\hM$ is
  less than~$M$, we do the opposite by proposing to move some rejections to
  after the last acceptance.
  
  When proposing additional rejections, we must also propose times for them
  among the current latent history.  There are~${\hM+N-1 \choose \hM-M}$
  such ways to insert these additional rejections into the existing latent
  history, such that the sampler terminates after the~$N$th acceptance.
  When removing rejections, we must choose which ones to place after the
  data, and there are~${M \choose M-\hM}$ possible sets.  Upon
  simplification, the proposal ratios for both addition and removal of
  rejections are identical:
  \begin{align*}
    \overbrace{
    \frac{
      q(M\!\leftarrow\!\hM)
      { \hM + N - 1 \choose \hM - M }
    }{
      q(\hM\! \leftarrow\! M)
      { \hM \choose \hM - M }
    }
    }^{\hM > M}
    =
    \overbrace{
    \frac{
      q(M\!\leftarrow\!\hM)
      { M \choose M - \hM }
    }{
      q(\hM\!\leftarrow\! M)
      { M + N - 1 \choose M - \hM }
    }
    }^{\hM < M}
    =
    \frac{
      q(M \!\leftarrow \!\hM)
      M! (\hM\!+\! N\! - \!1)!
    }{
      q(\hM\! \leftarrow\! M)
      \hM! (M\! +\! N\! -\! 1)!
    }.
  \end{align*}
  When inserting rejections, we propose the locations of the additional
  proposals, denoted~$\mcM^{+}$, and the corresponding values of the latent
  function, denoted~$\mcG^{+}_M$.  We generate~$\mcM^{+}$ by making~$\hM~-~M$
  independent draws from the base density.  We draw~$\mcG^{+}_M$ jointly
  from the Gaussian process prior, conditioned on all of the current latent
  state, i.e., $(\mcM,~\mcG_M,~\mcD,~\mcG_N)$. The joint probability of
  this state is
  \begin{multline}
    \label{eqn:insert-joint}
    p(\mcD, \mcM, \mcM^{+}, \mcG_N, \mcG_M, \mcG^{+}_M\given\theta,\psi)
    =
    \left[\prod_{n=1}^N \pi(\bx_n\given\psi) \; \Phi(g(\bx_n)) \right]\\
    \times
    \left[\prod_{m=1}^M \pi(\bx_m\given\psi) \; (1-\Phi(g(\bx_m)))\right]
    \left[\prod_{m=M+1}^{\hM} \!\!\!\! \pi(\bx_m\given\psi)\right]\\
    \times
    \distGP( \mcG_M, \mcG_N, \mcG^{+}_M \given \mcD, \mcM, \mcM^{+},
    \theta).
  \end{multline}
  The joint distribution in Equation~\ref{eqn:insert-joint} expresses the
  probability of all the base density draws, the values of the function
  draws from the Gaussian process, and the acceptance or rejection
  probabilities of the proposals \textit{excluding} the newly generated
  points.  When we make an insertion proposal, exchangeability allows us to
  shuffle the ordering without changing the probability; the only change is
  that now we must account for labeling the new points as rejections.  In
  the acceptance ratio, all terms except for the ``labeling probability''
  cancel.  The reverse proposal is similar, however we denote the removed
  proposal locations as~$\mcM^{-}$ and the corresponding function values
  as~$\mcG^{-}_M$.  The overall acceptance ratios for insertions or
  removals are
  \begin{align}
    \label{eqn:gpds-general-number}
    a_{\sf{hist-num}} = \begin{cases}
    \frac{
      q(M \leftarrow \hM) \; M! \; (\hM+N-1)!
    }{
      q(\hM \leftarrow M) \; \hM! \; (M+N-1)!
    }
    \prod_{g \in \mcG^{+}_M}(1-\Phi(g)) & \text{if $\hM > M$}\\
    \quad\\
    \frac{
      q(M \leftarrow \hM) \; M! \; (\hM+N-1)!
    }{
      q(\hM \leftarrow M) \; \hM! \; (M+N-1)!
    }
    \prod_{g \in \mcG^{-}_M}(1-\Phi(g))^{-1} & \text{if $\hM < M$}.
    \end{cases}
  \end{align}
  
  A simple and convenient way of implementing this procedure is to make
  limited proposals that either insert or delete only one latent rejection
  at a time.  We define a function~${\zeta(M,N):
    \naturals\times\naturals^{+}\to(0,1]}$ and propose inserting
    a new latent rejection with probability $\zeta$. Otherwise, with
    with probability $1-\zeta$, we propose removing a rejection.
    We must, of course, enforce ${\zeta(0,N) = 1}$.  With these limited
    proposals, the first case of Equation~\ref{eqn:gpds-general-number}
    (proposing one new latent rejection, i.e.,~${\hM=M+1}$) can be written
    as
  \begin{align}
  \label{eqn:gpds-insert}
  a_{\sf{hist-ins}} &= \frac{(1-\zeta(M+1,N))\;(M+N)\;
    (1-\Phi(g(\bx^{+})))}{\zeta(M,N)\;(M+1)} ,
  \end{align}
  where~$\bx^{+}$ is the proposed rejection location.  The
  location~$\bx^{+}$ is drawn from the base density~$\pi(\bx\given\psi)$.
  In the second case, if there is at least one latent rejection in the
  current history (${M>0}$), then the deletion of a single rejection is
  proposed, i.e.,~${\hM=M-1}$.  This deletion proposal has
  Metropolis--Hastings acceptance ratio
  \begin{align}
  \label{eqn:gpds-delete}
  a_{\sf{hist-del}} &=
  \frac{\zeta(M-1,N)\;M}{(1-\zeta(M,N))\;(M+N-1)\;(1-\Phi(g(\bx^{-})))},
  \end{align}
  where~$\bx^{-}$ is the location of the proposed removal.  The rejection
  to remove is chosen uniformly from among the~$M$ currently in the
  history.

  \subsubsection{Modifying latent rejection locations}
  Given the number of latent rejections~$M$ and the current latent
  function, we would like to sample from the locations of the rejections.
  Given the latent function, the locations of the rejections are
  independent.  We make perturbative proposals of new locations,
  conditionally sample the function from the Gaussian process and then
  accept or reject with Metropolis--Hastings.

  The current locations of the rejections are denoted~$\mcM$ and we draw a
  proposal~$\hmcM$ from a proposal distribution~$q(\hmcM\leftarrow\mcM)$.
  The values of the latent function at $\mcM$ are denoted $\mcG_M$ and we
  sample the function at $\hmcM$ jointly from the Gaussian process prior
  given $\mcD$, $\mcG_N$, $\mcM$, and $\mcG_M$.  The Metropolis--Hastings
  acceptance ratio of this proposal is
   \begin{align}
    a_{\sf{hist-locs}} &=
    \frac{
      q(\mcM \leftarrow \hmcM)
    }{
      q(\hmcM \leftarrow \mcM)
    }
      \prod^M_{m=1}
      \frac{\pi(\bhx_m\given\psi) \; (1-\Phi(\hg(\bx_m)))}
      {\pi(\bx_m\given\psi) \; (1-\Phi(g(\bx_m)))}.
  \end{align}

  \subsubsection{Modifying the latent function values}
  \label{sec:hist-latent-func}
  Conditioned on the number and location of the latent rejections, we must
  also sample from the latent function at both the data and rejection
  locations.  The conditional joint posterior distribution is
  \begin{multline}
    p(\mcG_N,\mcG_M \given\mcM,\mcD,\theta)
    = \distGP(\mcG_N,\mcG_M\given\mcD,\mcM,\theta) \\ \times
    \left[\prod^N_{n=1}\Phi(g(\bx_n))\right]
    \left[\prod^M_{m=1}\left(1-\Phi(g(\bx_m))\right)\right].
  \end{multline}
  This joint distribution is easily sampled using Hybrid (Hamiltonian)
  Monte Carlo~\citep{duane-etal-1987a}.  For numerical reasons we suggest
  performing gradient calculations in the ``whitened'' space resulting from
  applying the inverse Cholesky decomposition of the covariance matrix to
  the function values.

  Algorithm~\ref{alg:latent-history} implements the latent history
  algorithm in pseudocode, with the simple $q(\hM\leftarrow M)$ that
  proposes increasing or decreasing the number of latent rejections $M$ by
  one.

  \subsubsection{Hyperparameter inference}
  \label{sec:hist-hyper}
  Given a sample from the posterior on the latent history, we can also
  perform a Metropolis--Hastings step in the space of hyperparameters.  As
  in Section~\ref{sec:exchange-hyper}, we have hyperparameters~$\theta$ for
  the Gaussian process and~$\psi$ for the base density, with joint prior
  density~$p(\theta,\psi)$.  We introduce the proposal density
  $q(\htheta,~\hpsi~\leftarrow~\theta,~\psi)$ to make proposals $\htheta$
  and $\hpsi$.  The acceptance ratio for a Metropolis--Hastings step in the
  posterior of the hyperparameters, given the latent history, is
  \begin{multline}
    a_{\sf{hist-hp}} = \frac{
      q(\theta,\psi \!\leftarrow\! \htheta,\hpsi) \;
      p(\htheta,\hpsi) \;
      \distNorm(\{\mcG_M, \mcG_N\} \given \mcM, \mcD, \htheta)
    }{
      q(\htheta,\hpsi \!\leftarrow\! \theta,\psi) \;
      p(\theta,\psi) \;
      \distNorm(\{\mcG_M, \mcG_N\} \given \mcM, \mcD, \theta)
    }\\\times
    \left[
      \prod_{m=1}^M\frac{\pi(\bx_m\given\hpsi)}{\pi(\bx_m\given\psi)}
    \right]
    \left[
      \prod_{n=1}^N\frac{\pi(\bx_n\given\hpsi)}{\pi(\bx_n\given\psi)}
    \right].
  \end{multline}

  \begin{algorithm}[t]
    \begin{algorithmic}[1]

  \Ensure%
  \begin{minipage}[t]{\linewidth}
    \begin{itemize*}
    \item Number of MCMC iterations $R$
    \item Observed data $\mcD=\{\bx_n\}^N_{n=1}$
    \item Gaussian process covariance function $C(\bx,\bx'; \theta)$
    \item Base density $\pi(\bx\given\psi)$
    \item Location proposal density $q(\bhx_m\leftarrow\bx_m)$      
    \item Insert proposal probability function $\zeta(M,N)$
   \end{itemize*}
  \end{minipage}

  \Require%
  \begin{minipage}[t]{\linewidth}
    \begin{itemize*}
     \item $R$ samples of the latent history $\{ \mcM^{(r)}, \mcG^{(r)}_N,
      \mcG^{(r)}_M \}^R_{r=1}$
     \end{itemize*}
  \end{minipage}
  \vskip 0.25cm
  \State $\mcM \gets \emptyset$, $\mcG_M\gets\emptyset$
  \Comment{Start out with no latent rejections.}

  \State $\mcG_N\sim \distGP(g\given\mcD,\theta)$
  \Comment{Initialise the function at the data.}

  \For{$r \gets 1\ldots R$}
  \Comment{Take $R$ MCMC steps on the latent history.}

  \State $u_{\zeta} \sim \distUni(0,1)$
  \Comment{Draw a uniform random variate on $(0,1)$.}

  \If{$u_{\zeta} < \zeta(|\mcM|,N)$}
  \Comment{Decide whether to insert or delete.}

  \State $\bx^{+} \sim \pi(\bx\given\psi_{\pi})$
  \Comment{Draw a proposed rejection location.}

  \State $g(\bx^{+}) \sim
  \distGP(g\given\bx^{+},\mcD,\mcM,\mcG_M,\mcG_N,\theta)$
  \Comment{Draw the proposed function value.}

  \State $a_{\sf{hist-ins}} \gets \displaystyle
  \frac{(1-\zeta(|\mcM|+1,N))\;(|\mcM|+N)\;(1-\Phi(g(\bx^{+})))
  }{\zeta(|\mcM|,N)\;(|\mcM|+1)}$
  \Comment{Acceptance ratio.}

  \State $u_{\sf{ins}} \sim \distUni(0,1)$
  \Comment{Draw a uniform random variate on $(0,1)$.}
  \If{$u_{\sf{ins}} < a_{\sf{hist-ins}}$}
  \Comment{Metropolis--Hastings acceptance rule.}
  
  \State $\mcM\gets\mcM\cup\bx^{+}$, $\mcG_M\gets\mcG_M\cup g(\bx^{+})$
  \Comment{Add this new rejection.}

  \EndIf

  \ElsIf{$|\mcM|>0$}

  \State $m \sim \lceil \distUni(0,|\mcM|) \rceil$
  \Comment{Select one of the $M$ rejections at random.}

  \State $a_{\sf{hist-del}} = \displaystyle \frac{\zeta(|\mcM|-1,N)\;|\mcM|}{
    (1-\zeta(|\mcM|,N))\;(|\mcM|+N-1)\;(1-\Phi(g(\bx_m)))}$
  \Comment{Acceptance ratio.}

  \State $u_{\sf{del}} \sim \distUni(0,1)$
  \Comment{Draw a uniform random variate on $(0,1)$.}

  \If{$u_{\sf{del}} < a_{\sf{hist-del}}$}
  \Comment{Metropolis--Hastings acceptance rule.}

  \State $\mcM\gets\mcM\backslash\bx_m$, $\mcG_M\gets\mcG_M\backslash
  g(\bx_m)$
  \Comment{Remove the $m$th rejection.}
  \EndIf

  \EndIf

  \For{$m\gets 1\ldots M$}
  \Comment{Loop over the latent rejections.}
  
  \State $\bhx_m\sim q(\bhx_m\leftarrow\bx_m)$
  \Comment{Propose a new location.}
  \State $g(\bhx_m) \sim \distGP(g\given\bhx_m,\mcD,\mcM,\mcG_N,\mcG_M,\theta)$
  \Comment{Draw a function value from the GP.}

  \State $a_{\sf{hist-loc}}=\displaystyle\frac{
    q(\bx_m\leftarrow\bhx_m)\;\pi(\bhx_m)\;(1-\Phi(g(\bhx_m)))}{
    q(\bhx_m\leftarrow\bx_m)\;\pi(\bx_m)\;(1-\Phi(g(\bx_m)))}$
  \Comment{Acceptance ratio.}

  \State $u_{\sf{loc}} \sim \distUni(0,1)$
  \Comment{Draw a uniform random variate from $(0,1)$.}

  \If{$u_{\sf{loc}}<a_{\sf{hist-loc}}$}
  \Comment{Metropolis--Hastings acceptance rule.}
  
  \State $\bx_m \gets \bhx_m$, $g(\bx_m)\gets g(\bhx_m)$
  \Comment{Update the rejection.}

  \EndIf
  \EndFor

  \State $\mcG_N$,$\mcG_M \sim
  \textsf{HMC}(\mcG_N,\mcG_M\given\mcD,\mcM,\theta)$
  \Comment{Update function via Hybrid Monte Carlo.}

  \State $\mcM^{(r)}\gets\mcM$, $\mcG^{(r)}_N\gets\mcG_N$,
  $\mcG^{(r)}_M\gets\mcG_M$
  \Comment{Store the current version of the history.}

  \EndFor
  \State \Return $\{ \mcM^{(r)}, \mcG^{(r)}_N,
  \mcG^{(r)}_M \}^R_{r=1}$
  
\end{algorithmic}

    \caption{Simulate $R$ steps of a Markov chain on the latent history}
    \label{alg:latent-history}
  \end{algorithm}
  \afterpage{\clearpage}

  \subsubsection{Generating predictive samples}
  As with the exchange sampling approach in
  Section~\ref{sec:exchange-predictive}, it is possible to generate samples
  from the predictive density.  As each state in the Markov chain of the
  latent history inference is a rejection sampler state, it is simply a
  matter of continuing the rejection procedure forward to produce a new
  sample. 

  \subsection{Calculating the predictive density}
  We have shown that each inference method can yield predictive samples,
  but it is also natural to require that a density model provide an
  estimate of the normalized predictive density itself.  We use the
  method of~\citet{chib-jeliazkov-2001a}, which considers
  Metropolis--Hastings moves between a pair~$\bx$ and~$\bx'$. Using
  the base density~$\pi(\bx\given\psi)$ as the
  proposal density, the detailed balance condition for
  Metropolis--Hastings gives the identity
  \begin{multline}
    p(\bx, \bg, \theta,\psi)\;
    \pi(\bx'\given\psi)\;
    \min\left(1, \frac{\Phi(g(\bx'))}{\Phi(g(\bx))}\right)
    = \\ p(\bx', \bg, \theta, \psi)\;
    \pi(\bx\given\psi)\;
    \min\left(1, \frac{\Phi(g(\bx))}{\Phi(g(\bx'))}\right).
  \end{multline}
  We integrate both sides of this identity over~$x'$ and take the
  expectation of each side under the posterior over the function $\bg$ and
  the hyperparameters $\theta$ and $\psi$:
  \begin{multline*}
    \int\!\!\! \mathrm{d}\theta \!\!
    \int\!\!\! \mathrm{d}\psi \!
    \int\!\!\! \mathrm{d}\bg \;
    p(\bg,\theta,\psi\given\mcD)
    \int\!\! \mathrm{d}x'\;
    p(\bx \given \bg, \theta,\psi)\;
    \pi(\bx'\given\psi)\;
    \min\left(1, \frac{\Phi(g(\bx'))}{\Phi(g(\bx))}\right)
    = \\ 
    \int\!\!\! \mathrm{d}\theta \!
    \int\!\!\! \mathrm{d}\psi \!
    \int\!\!\! \mathrm{d}\bg\;
    p(\bg,\theta,\psi\given\mcD)
    \int\!\! \mathrm{d}x'\;
    p(\bx' \given \bg, \theta, \psi)\;
    \pi(\bx\given\psi)\;
    \min\left(1, \frac{\Phi(g(\bx))}{\Phi(g(\bx'))}\right).
  \end{multline*}
  We observe that
  \begin{align*}
    p(\bg,\theta,\psi\given\mcD)\;
    p(\bx \given \bg, \theta,\psi)
    = p(\bx, \bg,\theta,\psi\given\mcD)
    = p(\bx\given\mcD)\;p(\bg,\theta,\psi\given x,\mcD)
  \end{align*}
  and so we may find the predictive density via
  \begin{align}
    \label{eqn:predictive-ratio}
    p(\bx\given\mcD) &= \frac{
    \int\! \mathrm{d}\theta \!
    \int\! \mathrm{d}\psi \!
    \int\! \mathrm{d}\bg \!
    \int\! \mathrm{d}x' \;
    p(\theta,\psi,\bg,x'\given\mcD)\;
    \pi(\bx\given\psi)\;\min\left(1, \frac{\Phi(g(\bx))}{\Phi(g(\bx'))}\right)}
    {
    \int\! \mathrm{d}\theta \!
    \int\! \mathrm{d}\psi \!
    \int\! \mathrm{d}\bg \!
    \int\! \mathrm{d}x' \;
    p(\theta,\psi,\bg\given x, \mcD)\;
    \pi(\bx'\given\psi)\;\min\left(1,
    \frac{\Phi(g(\bx'))}{\Phi(g(\bx))}\right)
    }      
  \end{align}
  Both the numerator and the denominator in
  Equation~\ref{eqn:predictive-ratio} are expectations.  The top is an
  expectation under the posterior and the bottom is an expectation under
  the posterior where the data has been augmented with~$x$:
  \begin{align}
    p(\bx\given\mcD)
    &= \frac{\mathbb{E}_{p(\bg,\theta,\psi,x'|\mcD)}
      \left[\pi(\bx\given\psi)\; \min\left(1,
      \frac{\Phi(g(\bx))}{\Phi(g(\bx'))}\right)\right]}
    {\mathbb{E}_{p(\bg,\theta,\psi|\mcD,x)}
      \left[
      \mathbb{E}_{\pi(\bx'\given\psi)}
      \left[\min\left(1,
          \frac{\Phi(g(\bx'))}{\Phi(g(\bx))}\right)\right]\right]}.
  \end{align}
  The numerator can be estimated directly as part of the MCMC inference.
  After each Markov step, generate a predictive sample~$x'$ and record the
  transition probabilities.  The denominator requires a Markov chain to be
  run with a data set augmented by the predictive location~$x$.  At each
  step in the Markov chain, a sample $x'$ is generated from the base
  density and the transition probabilities are evaluated.
  
  \section{Examples}

  \begin{figure}[t]
    \centering%
    \subfloat[][True density, data and histogram]{%
      \centering%
      \includegraphics[width=0.49\textwidth]{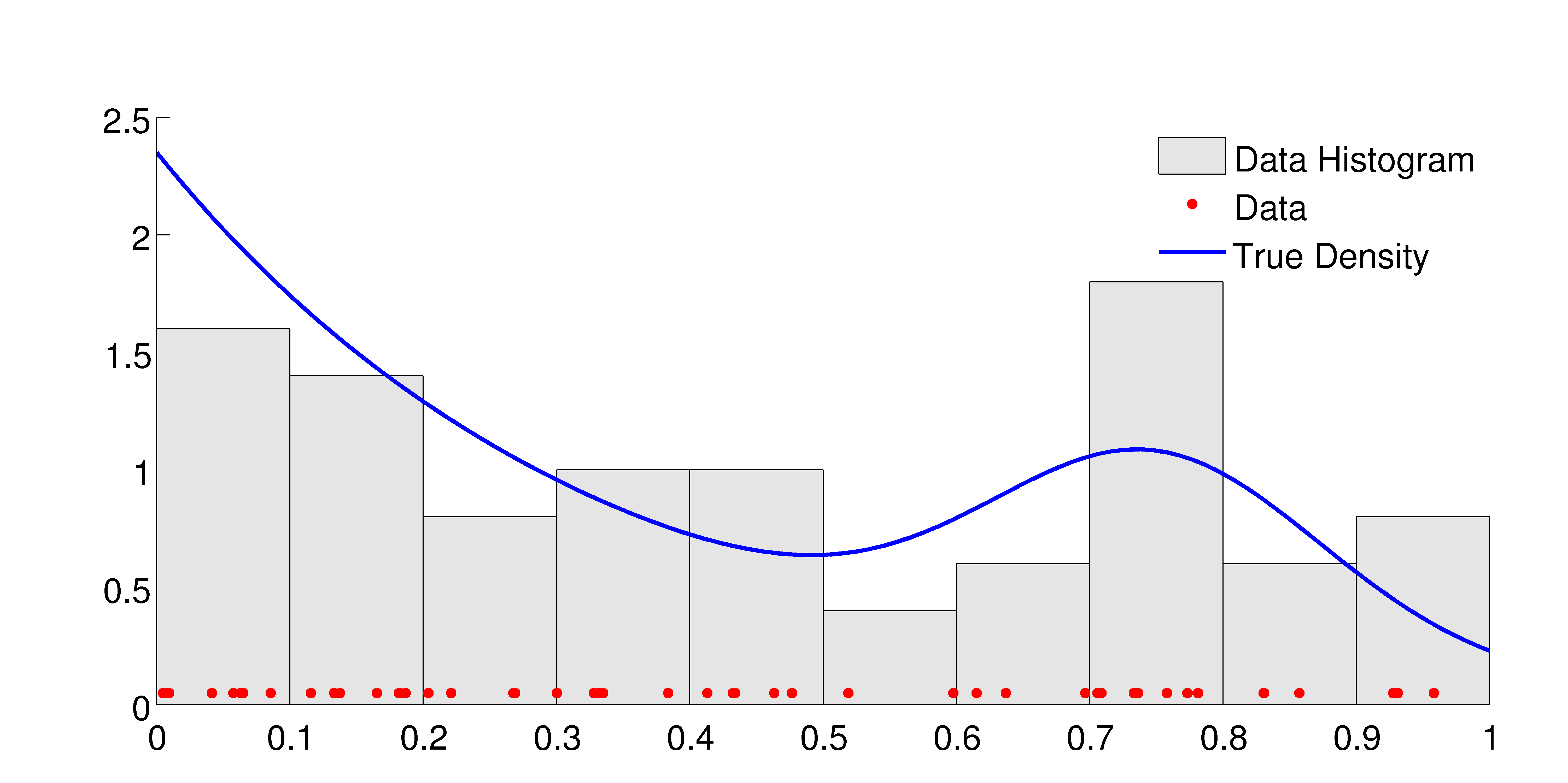}%
      \label{fig:lenkmix:data}%
    }~%
    \subfloat[][Density estimates]{%
      \centering%
      \includegraphics[width=0.49\textwidth]{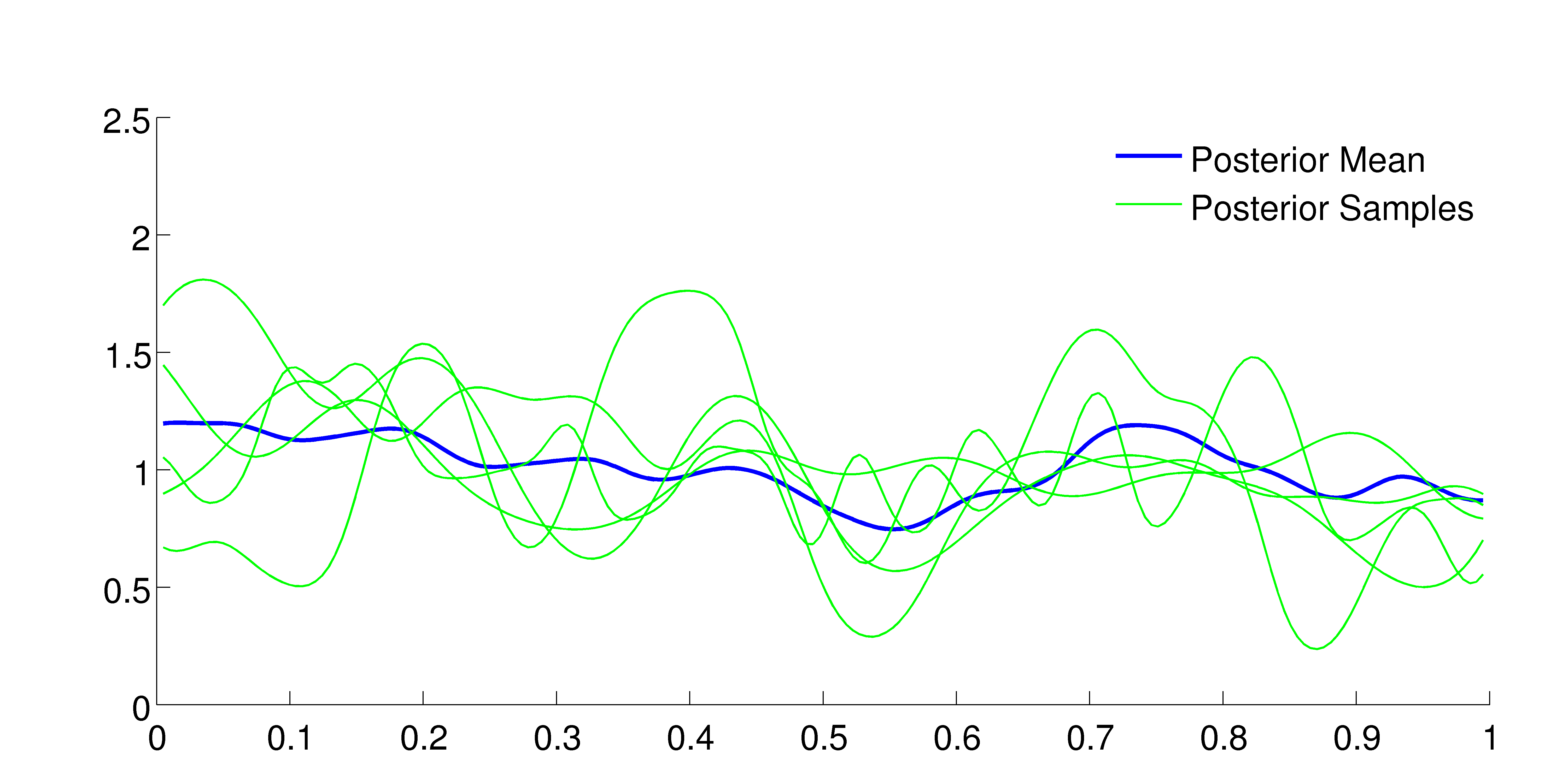}%
      \label{fig:lenkmix:gpds}%
    }\\%
    \subfloat[][Histogram of rejection locations]{%
      \centering%
      \includegraphics[width=0.49\textwidth]{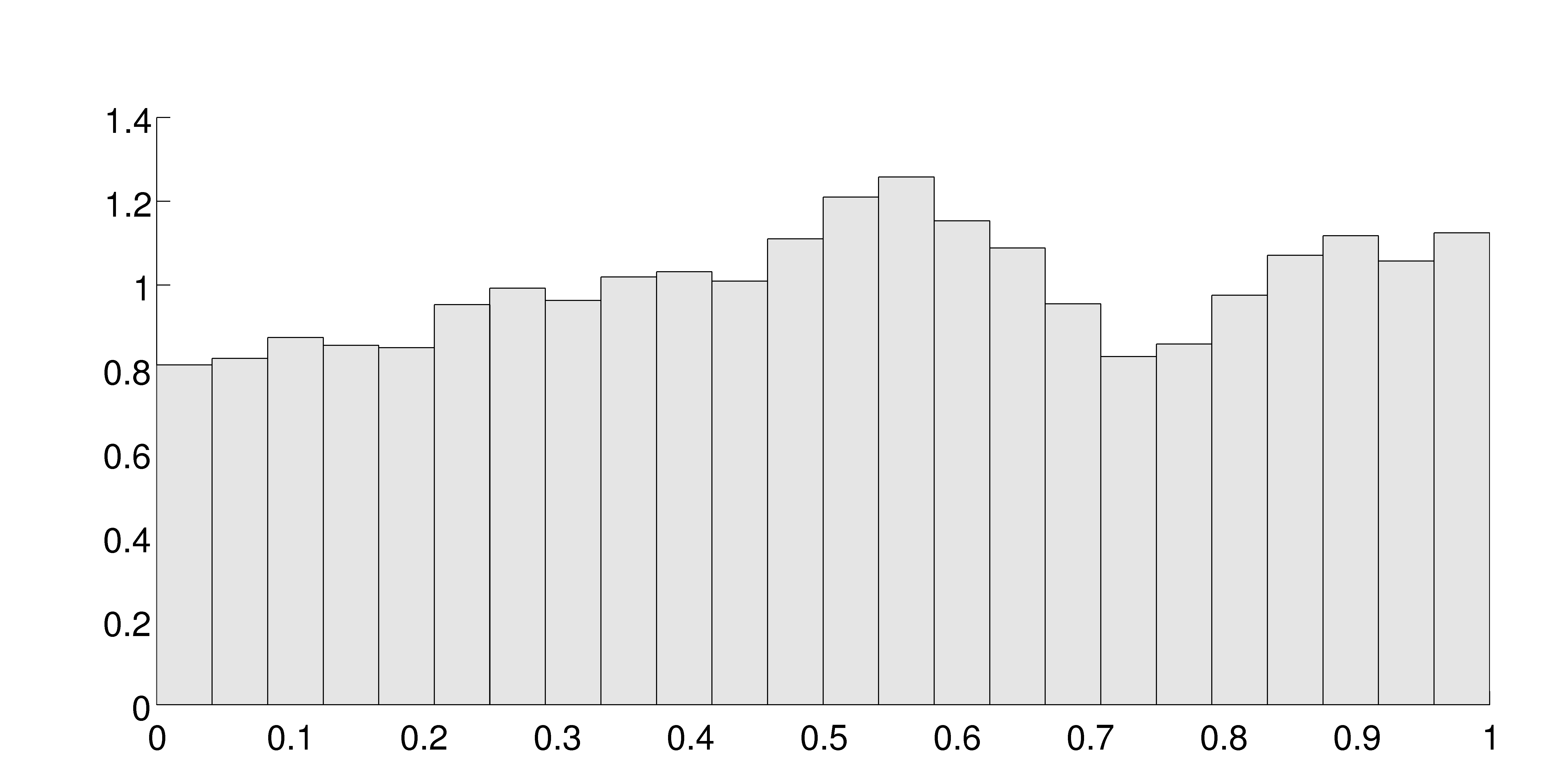}%
      \label{fig:lenkmix:rej-locs}%
     }~%
    \subfloat[][Histogram of number of rejections, $M$]{%
      \centering%
      \includegraphics[width=0.49\textwidth]{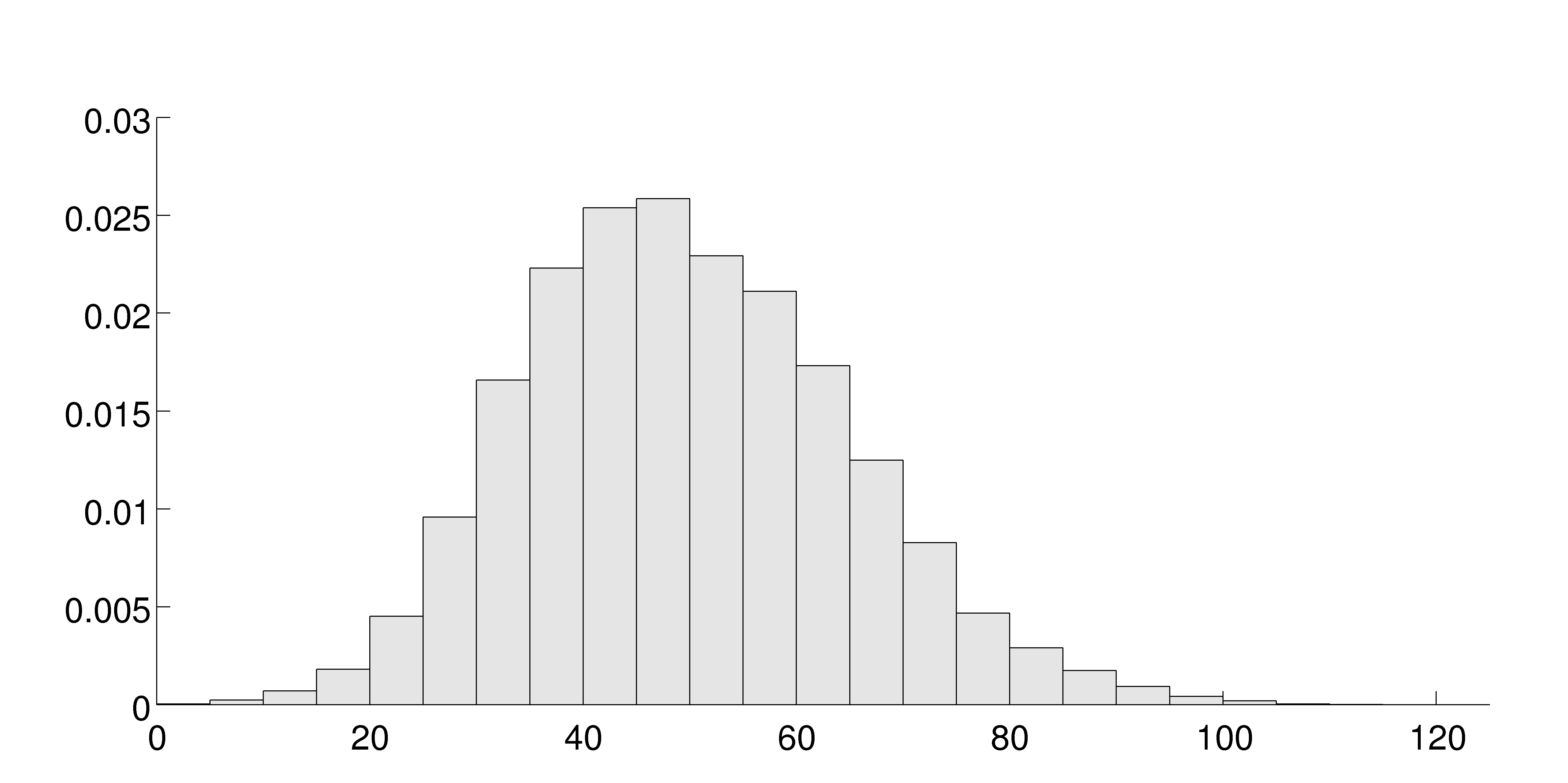}%
      \label{fig:lenkmix:rej-num}%
    }    
    \caption{Bounded one-dimensional example}
    \label{fig:lenkmix}
\end{figure}

  \subsection{One-Dimensional Bounded Density}
  We examined the GPDS on the one-dimensional problem studied by
  \citet{lenk-1991a} and \citet{tokdar-2007a}.  It is a mixture of an
  exponential and normal density on $[0,1]$:
  \begin{align}
    \label{eqn:lenkmix}
    f_1(x) &= \frac{3}{4} \cdot 3\exp\{-3x\}
    + \frac{1}{4}\cdot\left(\frac{\pi}{32}\right)^{-\half}
    \exp\Bigg\{-32\left(x-\frac{3}{4}\right)^2\Bigg\}.
  \end{align}
  50 independent observations were drawn from this density and latent
  history inference with the GPDS was applied to it.
  Figure~\ref{fig:lenkmix:data} shows a histogram of the observations and
  the true density.  The base density for the GPDS was chosen to be the
  uniform distribution on~$(0,1)$.  The covariance function used was the
  stationary squared exponential:
  \begin{align}
    \label{eqn:squared-exponential}
    C(x,x') &=
    \alpha^2\exp\left\{-\half\left(\frac{x-x'}{\ell}\right)^2\right\}
  \end{align}
  and the parameters~$\alpha$ and~$\ell$ were included in MCMC sampling for
  both the GPDS and the logistic Gaussian process.  The priors used for the
  Gaussian process hyperparameters were
  \begin{align}
    \label{eqn:lenkmix:amp-prior}
    \ln\alpha &\sim \distNorm(\mu=1,\, \sigma=0.5)\\
    \label{eqn:lenkmix:ls-prior}
    \ln\ell   &\sim \distNorm(\mu=0.05,\, \sigma=0.5).
  \end{align}
  The Markov chain was simulated for 50,000 iterations, with the first
  10,000 discarded as burn-in.  Figure~\ref{fig:lenkmix:gpds} shows the
  predictive density from the MCMC run, along with several posterior
  samples.  Figure~\ref{fig:lenkmix:rej-locs} shows a histogram of the
  locations of the rejections in the latent history inference, and
  Figure~\ref{fig:lenkmix:rej-num} is a histogram of the number of
  rejections in samples from the latent history.

  \subsection{Two-Dimensional Unbounded Density}
\begin{figure}[t]
    \centering%
    \subfloat[][True density and data]{%
      \centering%
      \includegraphics[width=0.49\textwidth]{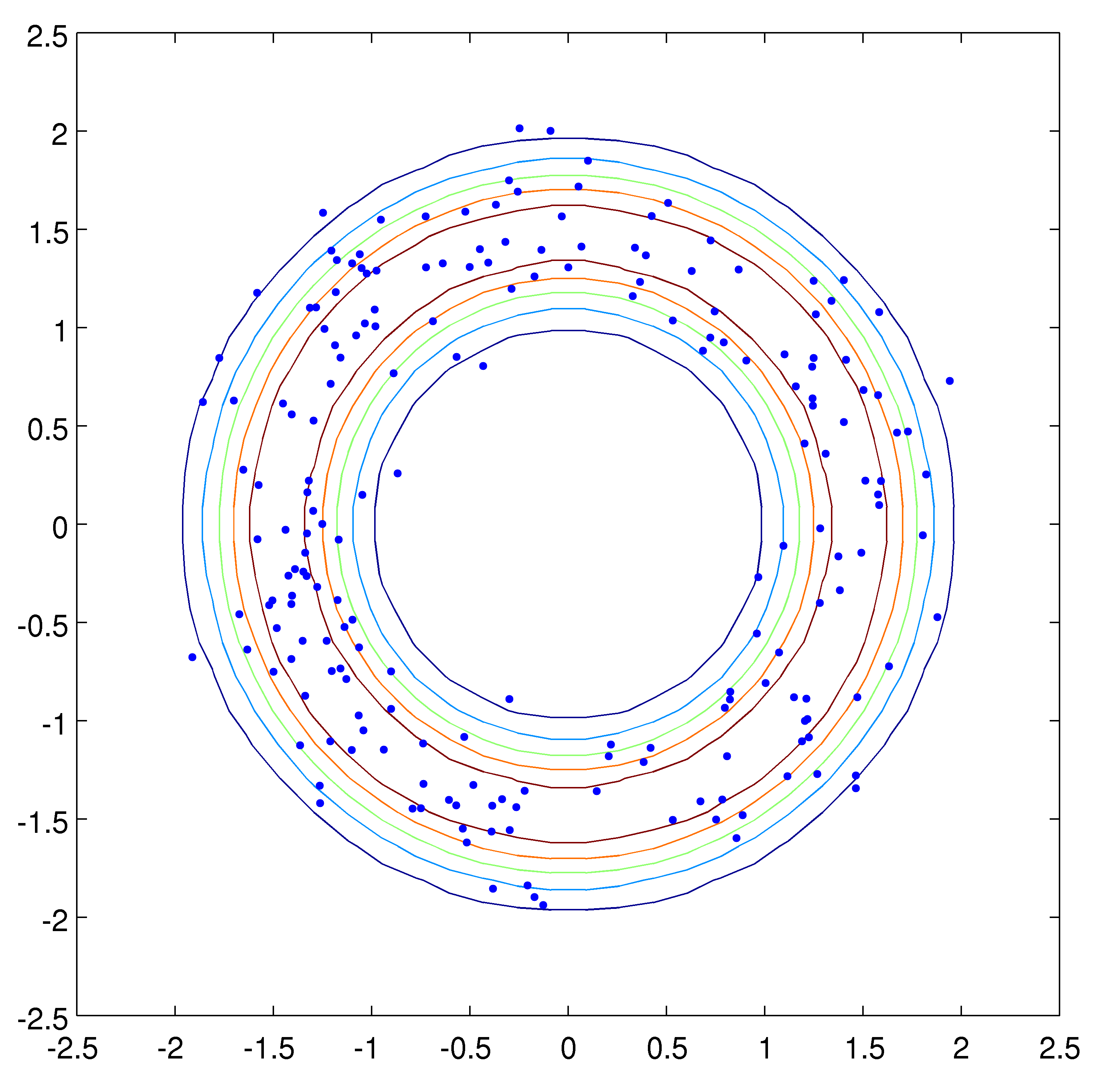}%
      \label{fig:ring-true}
    }~%
    \subfloat[][Density estimate and rejection snapshot]{%
      \centering%
      \includegraphics[width=0.49\textwidth]{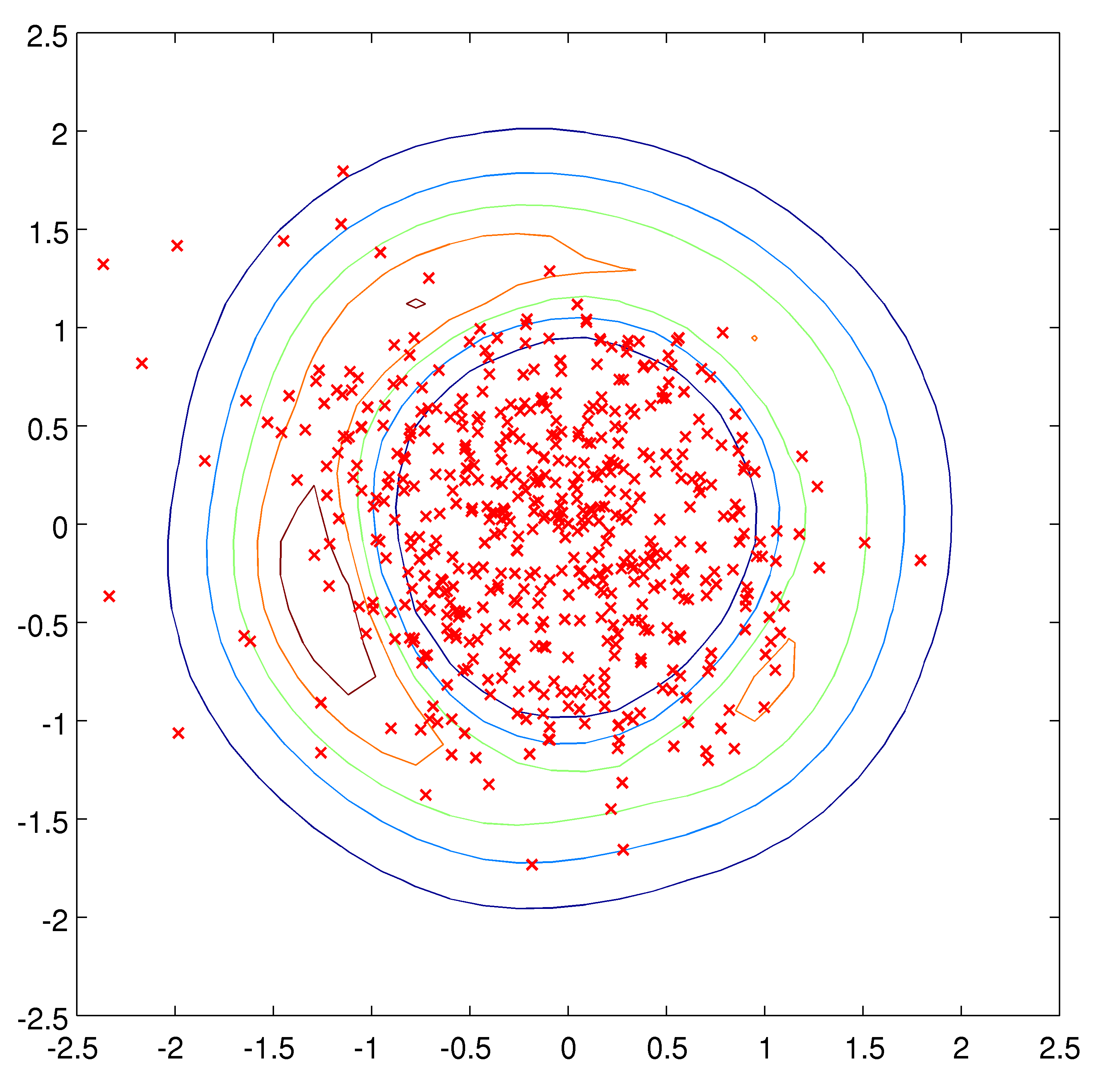}%
      \label{fig:ring-mean}
    }
    \caption{Synthetic ring mixture example}
   \end{figure}

  We generated 200 independent observations from a two-dimensional location
  mixture of Gaussians, where the means are drawn uniformly from a ring of
  radius~$3/2$, centred at the origin.  The Gaussians have a variance
  of~$1/16$ so that the data density is
  \begin{align*}
    f_2(x_1,x_2) &=  \frac{4}{9\pi}\int_{-\pi}^{\pi}\!\!\!\!\mathrm{d}\vartheta\;
    \distNorm(x_1\,;\,3/2\cos \vartheta,\sigma=1/4)
    \distNorm(x_2\,;\,3/2\sin \vartheta,\sigma=1/4).
  \end{align*}
  We used the two-dimensional isotropic variant of the covariance function
  given by Equation~\ref{eqn:squared-exponential} and used a Gaussian
  distribution for the base density, inferring the mean and covariance as
  in Section~\ref{sec:hist-hyper}.  We simulated the Markov chain for
  50,000 iterations, discarding 10,000 as burn-in.  The true density and
  the observed data are shown in Figure~\ref{fig:ring-true}, while the
  posterior predictive density and a posterior sample of rejection location
  are shown in Figure~\ref{fig:ring-mean}.  As expected, the rejections
  tend to accumulate in the center of the ring, where the base density
  places mass but the predictive density should be low.
   
  \section{Discussion}

  \subsection{Computational issues}
  Computation with a Gaussian process is expensive.  If the GP is realised
  on~$R$ points, the space complexity of storing the covariance (Gram)
  matrix is~$O(R^2)$ and the time complexity of decomposing (or inverting)
  the matrix is~$O(R^3)$.  The time cost of this decomposition will be the
  asymptotically-dominating factor when performing GPDS inference using
  either exchange sampling or the latent history method.

  \subsection{Comparing exchange sampling and latent history inference}
  Modeling of probability densities is fundamentally different from
  regression.  In regression and classification, one conditions on having
  seen data in the input space when performing inference and prediction.
  In these cases, it is necessary only to model the function at places
  where data have been observed, or at predictive query locations.  In
  density modeling, however, the places with low density are just as
  important to the model as those with high density.  Unfortunately, it is
  unlikely to have observed data in regions with low density, so a
  representation of the function only at locations where there are data is
  not adequate for the inference we wish to perform.  One might think of
  defining a density as analogous to putting up a tent: pinning the canvas
  down with pegs (or stakes) is just as important as putting up poles.  In
  exchange sampling, the ``pegs'' are inferred implicitly as rejections
  along the way to generating fantasy data.  At each exchange sampling
  step, a new tent is constructed --- complete with its own pegs --- and
  asked to explain the data.  In the latent history model, however, the
  tent is modified one piece at a time: pegs and poles are inserted,
  removed, and adjusted gradually to explain the data.

  It is possible also to see that the latent history model is likely to
  require fewer samples from the Gaussian process as it proceeds.  Consider
  the Gaussian process density sampler: when the latent history method is
  at equilibrium, its state will have some typical number of latent
  rejections~$M$.  This is about the same number of rejections as would be
  expected to occur during an exchange sampling fantasy.  However, to find
  the acceptance ratio in exchange sampling it is \textit{also} necessary
  to evaluate against the observed data after fantasising.  This means that
  the Gaussian process in exchange sampling requires at least~$2N+M$
  evaluations to make a Metropolis--Hastings move, while the latent history
  method requires only~$N+M$.  This does not even consider the expansion of
  state that occurs when exchange sampling rejects a proposal, and
  additional fantasy data are incorporated into the Markov state.  As the
  time complexity of computation in the Gaussian process grows cubically in
  the number of data, exchange sampling can become rapidly more expensive.

  Another reason that the latent history method is preferable to exchange
  sampling is that it requires less bookkeeping about the
  function~$g(\bx)$.  The state of the exchange sampling Markov chain is
  the uncountably-infinite object~$g(\bx)$.  The innovation of the method
  is that through retrospective sampling we are able to make
  Metropolis--Hastings moves with only a finite number of computations.
  This retrospective sampling, however, means that information discovered
  about a particular~$g(\bx)$ must be retained for as long as that function
  is relevant to the current Markov state.  In contrast, the state of the
  Markov chain when performing latent history inference only
  includes~$g(\bx)$ at the latent rejections or thinned events.  That is,
  rather than an uncountably-infinite object~$g(\bx)$, the Gaussian process
  in the latent history model conditions on a finite set of points in the
  input space.  This means that the values of the function do not need to
  be kept in memory, except for at the data and at the locations of the
  rejections or thinned events.  This contrast can also be seen in the
  difference between the joint distributions that describe the two
  inference methods for the Gaussian process density sampler.  In exchange
  sampling, when writing Equation~\ref{eqn:simple-exchange-joint}, we
  use~$\bg$ to denote~$g(\bx)$ as an infinite vector.  When writing the
  posterior distribution on the latent history, however, we do not need to
  denote an infinite function.  Equation~\ref{eqn:gpds-lh-joint} only
  defines a distribution on the function values at the data and the latent
  rejections.
  
  Finally, while the latent history method enables efficient Hamiltonian
  Monte Carlo sampling of the latent function values, it is not clear how
  to combine HMC with exchange sampling.
  
  \subsection{Restricting the function space}
  With both the exchange sampling and latent history methods, incorporating
  fewer latent rejections (``tent pegs'') into the Gaussian process results
  in improved efficiency.  For a given~$g(\bx)$, the expected number of
  rejections is~$N(\mcZ_{\pi}[g]^{-1}-1)$.  This expression is derived from
  the observation that~$\pi(\bx\given\psi)$ provides an upper bound on the
  function~$\Phi(g(\bx))\,\pi(\bx\given\psi)$ and the ratio of acceptances to
  rejections is determined by the proportion of the mass
  of~$\pi(\bx\given\psi)$ contained by~$\Phi(g(\bx))\,\pi(\bx\given\psi)$.  One
  problem with inference is that there are many functions $g(\bx)$ that can
  explain the data equivalently, as~$\Phi(g(\bx))\,\pi(\bx\given\psi)$ is
  unnormalised.  Many of these~$g(\bx)$ will cause~$\Phi(g(\bx))$ to be close
  to zero, resulting in many rejections.  The Gaussian process prior might
  only provide weak regularisation to prevent this.

  One way to improve this situation is to require that the
  function~$g(\bx)$ be pinned to zero for some~$\bx_0$.  This
  prevents~$\Phi(g(\bx))\,\pi(\bx\given\psi)$ from being small everywhere
  and reduces the redundancy in the prior that occurs due to normalisation.
  We use the base density~$\pi(\bx\given\psi)$ as a prior on~$\bx_0$ and
  treat it as a hyperparameter for the Gaussian process.  We can then use
  the inference methods of Sections~\ref{sec:exchange-hyper}
  and~\ref{sec:hist-hyper} to infer an appropriate~$\bx_0$.
  
  \subsection{The logistic Gaussian process}
  The Gaussian process is an appealing prior on functions due to the
  ability to specify the smoothness and differentiability properties of
  sample realizations via a covariance function, without choosing an
  explicit set of basis functions.  This flexibility and intuition has led
  to interest in applying Gaussian processes to density modeling via the
  logistic Gaussian process introduced by \citet{leonard-1978a} and further
  developed by \citet{lenk-1988a,lenk-1991a}.  If~$g(\bx)$ is a random
  function drawn from a Gaussian process, then the logistic GP arrives at a
  density~$f(\bx)$ on a closed interval~$\mcI$ via
  \begin{align}
    \label{eqn:logistic-gp}
    f(\bx) &= \frac{ e^{g(\bx)} }{\int_{\mcI}e^{g(\bx)}}
  \end{align}
  for $\bx \in \mcI$.  On a bounded interval and with minor constraints on
  the covariance function, the integral in the denominator exists and the
  density is well-defined~\cite{tokdar-2007a}.  The distribution on
  densities is closed under Bayesian updating.  As in
  Equation~\ref{eqn:transformation}, however, it is generally impossible to
  integrate an infinite-dimensional random function and so likelihood-based
  calculations are intractable.  The GP-based prior we have presented in
  this paper allows exact inference computation despite this intractability
  by constructing a generative model, but no such method is known for the
  logistic Gaussian process.

  In order to perform inference with the logistic Gaussian process, several
  finite-dimensional approximations have been introduced.
  \citet{lenk-1991a,lenk-2003a} proposes an approximation of the logistic
  GP by a truncated Karhunen--Lo{\`e}ve expansion evaluated on a grid.
  \citet{tokdar-2007a} uses a finite-dimensional approximation to the
  logistic Gaussian process by parameterizing the function values on a grid
  and then imputing other values from the conditional mean.  The
  normalisation constant is estimated via a numeric method, e.g.,the
  trapezoidal rule or Simpson's rule.

  The approach of expanding the density as a finite Fourier series, as
  described by \citet{lenk-2003a}, is appealing in a single dimension as
  one can parameterize the function in terms of coefficients with
  independent Gaussian priors.  The variances of these priors arise
  directly from the Gaussian process covariance function.  As noted by
  \citet{lenk-2003a}, however, the number of Fourier coefficients required
  grows exponentially with dimension.  Finding higher-dimensional bases
  that are rich enough to express interesting structure while also allowing
  efficient computation is considered an open problem.

  The imputation method of \citet{tokdar-2007a} extends to the multivariate
  case more straightforwardly.  While a lattice does not scale well to many
  dimensions, the imputation approximation does not necessarily require a
  grid.  \citet{tokdar-2007a} proposes a method of inferring appropriate
  knot locations and explores this on a two-dimensional test problem using
  reversible jump Markov chain Monte Carlo \citep{green-1995a}.  This has a
  similar motivation to the model presented in this paper: adapt the
  parameterization of the Gaussian process as the data demands.  The GPDS
  achieves this via a fully-nonparametric generative model,
  \citet{tokdar-2007a} specifies a finite-dimensional surrogate model with
  the dimensionality selected as a part of inference.  Additionally, it is
  unclear in \citet{tokdar-2007a} how the normalization constant is to be
  effectively estimated when the knots are irregularly arranged.  It is
  suggested to perform imputation to a grid from the known knots, but this
  reintroduces some aspects of the problems of lattices in high dimensions.
  In contrast, the GPDS inference discussed in the present paper explicitly
  avoids these problems by performing computation without
  evaluating~$\mcZ_{\pi}[\bg]$.

  \section*{Acknowledgements}
  The authors wish to thank Radford Neal and Zoubin Ghahramani for valuable
  comments.

  \bibliographystyle{abbrvnat}
  \bibliography{draft}

\end{document}